%% file: fkr1_v2.tex
\documentclass[aps,nofootinbib,floatfix,amsmath,amssymb,preprint,superscriptaddress,groupedaddress]{revtex4-1}
\pdfoutput=1  

\usepackage{hyperref}

\usepackage{rotating}

\usepackage[font=small,labelfont=bf]{caption}
\usepackage{subfig}

\usepackage{feynmp}
\usepackage{dcolumn}
\usepackage{array}
\usepackage[english]{babel}

\usepackage{xspace}

\usepackage{bm}

\input{fkr1macros}

\newcommand{\mxnewcommand}[2]{\newcommand{#1}{\ensuremath{#2}\xspace}}
\newcommand{\mxrenewcommand}[2]{\renewcommand{#1}{\ensuremath{#2}\xspace}}

\newcommand{\xnewcommand}[2]{\newcommand{#1}{#2\xspace}}
\newcommand{\xrenewcommand}[2]{\renewcommand{#1}{#2\xspace}}

\newcommand{\roots}[1]{\ensuremath{\sqrt{s}={#1}\,\text{TeV}}\xspace}

\xnewcommand{\lhc}{\texttt{LHC}}
\xnewcommand{\higgs}{\texttt{Higgs}}
\xnewcommand{\planck}{\texttt{Planck}}

\renewcommand{\gev}{\ensuremath{\,\text{GeV}}\xspace}
\mxrenewcommand{\tev}{\,\text{TeV}}

\mxrenewcommand{\mhalf}{m_{1/2}}
\mxrenewcommand{\mzero}{m_0}
\mxrenewcommand{\azero}{A_0}
\mxrenewcommand{\tanb}{\tan\beta}
\mxrenewcommand{\sgnmu}{\text{sign}\,\mu}
\mxnewcommand{\invalpha}{1/\alphaemmz}
\mxrenewcommand{\alphas}{\alphasmzms}

\renewcommand{\(}{\left(}
\renewcommand{\)}{\right)}
\renewcommand{\[}{\left[}
\renewcommand{\]}{\right]}

\newcommand{\priorf}[1]{\ensuremath{\pi\(#1\)}}
\newcommand{\likef}[1]{\ensuremath{\mathcal{L}\(#1\)}}
\newcommand{\gaussian}[3]{\ensuremath{\exp \[-\frac{\(#1 - #2\)^2}{2#3} \]}}
\newcommand{\probg}[2]{\ensuremath{p\(#1|#2\)}}

\mxrenewcommand{\like}{\mathcal{L}}
\mxnewcommand{\prior}{\pi}
\mxnewcommand{\params}{m}
\mxrenewcommand{\ev}{\mathcal{Z}}

\newcommand{\s}[1]{\ensuremath{\tilde{#1}}}
\newcommand{\neut}[1]{\ensuremath{{\chi}^0_{#1}}\xspace}
\newcommand{\charg}[1]{\ensuremath{{\chi}^\pm_{#1}}\xspace}

\mxnewcommand{\mi}{m_{\neut{1}}}
\mxnewcommand{\mii}{m_{\s{\ell}}}
\mxnewcommand{\miii}{m_{\neut{2}}}
\mxnewcommand{\miv}{m_{\s{q}}}
\mxnewcommand{\hmi}{\hat{e}_{\mi}}
\mxnewcommand{\hmii}{\hat{e}_{\mii}}
\mxnewcommand{\hmiii}{\hat{e}_{\miii}}
\mxnewcommand{\hmiv}{\hat{e}_{\miv}}

\mxnewcommand{\pmm}{(\mzero,\,\mhalf)}
\mxnewcommand{\pat}{(\azero,\,\tanb)}
\mxnewcommand{\pcs}{(\mi,\,\sigsip)}
\xrenewcommand{\stauc}{stau-coannihilation}

\let\brbsmumuold\brbsmumu
\xrenewcommand{\brbsmumu}{\brbsmumuold}
\let\sigsipold\sigsip
\xrenewcommand{\sigsip}{\sigsipold}
\let\mtold\mt
\xrenewcommand{\mt}{\mtold}
\let\egold\eg
\xrenewcommand{\eg}{\egold}

\usepackage{color}

\begin{document}

\title{Reconstructing CMSSM parameters at the LHC with \roots{14} via
  the golden decay channel}

\author{Andrew Fowlie}
\email{Andrew.Fowlie@KBFI.ee}
\affiliation{National Institute of Chemical Physics and Biophysics, Ravala 10,
Tallinn 10143, Estonia}

\author{Malgorzata Kazana}
\email{Malgorzata.Kazana@fuw.edu.pl}
\affiliation{National Centre for Nuclear Research,
  Ho{\. z}a 69, 00-681 Warsaw, Poland}

\author{Leszek Roszkowski}
\email{L.Roszkowski@sheffield.ac.uk}
\altaffiliation{On leave of absence from Department of Physics and Astronomy, University of
  Sheffield, Sheffield S3 7RH, England.}
\affiliation{National Centre for Nuclear Research, Ho{\. z}a 69, 00-681 Warsaw, Poland}

\date{\today}

\begin{abstract}
  We identify a benchmark point in the CMSSM's heavy \stauc region,
  which is favored by experiments, and demonstrate that it could be
  accessible to the LHC at \roots{14} with
  $300\invfb$ of integrated luminosity via a golden decay measurement.  With
  Monte-Carlo, we simulate sparticle production and subsequent golden
  decay at the event level and perform pseudo-measurements of
  sparticle masses from kinematic endpoints in invariant mass
  distributions. We find that two lightest neutralino masses and the first and
  second generation left-handed slepton and squark masses could be rather
  precisely measured with correlated uncertainties. We investigate
  whether from such measurements one could determine the CMSSM's
  Lagrangian parameters by including a likelihood from our
  pseudo-measurements of sparticle masses in a Bayesian analysis of
  the CMSSM's parameter space. We find that the CMSSM's parameters can
  be accurately determined, with the exception of the common
  trilinear parameter. Experimental measurements of the relic density by Planck
  and the Higgs boson's mass slightly improve this determination,
  especially for the common trilinear parameter. Finally, within our benchmark
  scenario, we show that the neutralino dark matter will be accessible
  to direct searches in future one tonne detectors.
\end{abstract}

\maketitle

\section{\label{sec:intro}Introduction}
Despite the absence of softly-broken supersymmetry (SUSY) in searches at the Large Hadron
Collider (LHC)\cite{Aad:2012fqa,Chatrchyan:2012uea}, SUSY remains a leading and
attractive candidate for new physics beyond the Standard Model (SM).

The discovery by both ATLAS\cite{Chatrchyan:2012ufa} and
CMS\cite{Aad:2012tfa} of a Higgs-like boson with mass $\mh \simeq 126\gev$
indicates that the SUSY breaking scale is likely to be rather high,
exceeding $1\tev$\cite{Roszkowski:2012uf,Kowalska:2013hha,Baer:2012mv,Buchmueller:2012hv}.
This, on the one hand, weakens naturalness
arguments for SUSY but, on the other, is consistent with stringent
bounds on superpartner masses from the LHC, and also is consistent
with the fact that superpartners have not been observed yet, either directly or via loop
effects in rare flavor changing processes.

In fact, within the framework of grand-unified (or, in other words, GUT-constrained)
SUSY models, like the popular Constrained Minimal Supersymmetric
Standard Model (CMSSM)\cite{Kane:1993td},\footnote{The CMSSM imposes a
    simple pattern on the MSSM soft-breaking parameters at the GUT
    scale, in which gaugino masses are unified to \mhalf, scalar masses
    are unified to \mzero and trilinear couplings are unified to
    \azero. The superpotential bilinear $\mu$ and the soft-breaking
    bilinear $B$ are traded via electroweak symmetry breaking (EWSB)
    conditions for $\tanb = v_u/v_d$ and $\mz$, while \sgnmu\ remains
    undetermined. Whilst the CMSSM posits no particular SUSY breaking mechanism, it is
    phenomenologically similar to minimal supergravity\cite{Arnowitt:1992aq,Chamseddine:1982jx}.}
the rather large Higgs boson mass favours a region of parameter space in which
the lightest neutralino is higgsino-like with mass close to
$1\tev$\cite{Strege:2012bt,Cabrera:2012vu,Kowalska:2013hha}. The same
is true in the Non-Universal Higgs Model (NUHM)\cite{Roszkowski:2009sm,Strege:2012bt,Cabrera:2012vu}.
This new high \msusy\ region appears at multi-TeV ranges of \mzero
and \mhalf, in addition to the previously
identified \stauc (SC), $\ha$-funnel (AF) and focus point/hyperbolic
branch (FP/HB) regions.
The favored regions are primarily determined by the relic density of the lightest
neutralino, which is assumed to be dark matter.

\renewcommand{\stauc}{SC\xspace}

Direct lower limits on SUSY masses from ATLAS and CMS have, at small
$\mzero\lsim400\gev$, pushed up the scalar mass parameter to
$\mhalf\gsim850\gev$, while at multi-TeV \mzero, the bound is much
weaker, $\mhalf\gsim500\gev$\cite{ATLAS-CONF-2013-047,Aad:2014wea}. Measurements
of the rare decay \brbsmumu\ at LHCb\cite{Aaij:2013aka} and CMS\cite{Chatrchyan:2013bka},
which show no convincing deviations from the SM value, within
constrained SUSY imply that
the pseudoscalar Higgs
boson, \ha, is heavy and also push the AF region up to $\mhalf\gsim1\tev$.
In the FP/HB region at large \mzero\ one struggles to reproduce
$\mh\sim125\gev$\cite{Kowalska:2013hha},
and is now severely constrained\cite{Roszkowski:2014wqa}
by new upper limits on the spin-independent WIMP-proton
scattering cross section, \sigsip,  relevant to the direct detection of dark
matter, from the LUX experiment\cite{Akerib:2012ys,Akerib:2013tjd}.

Although the new $\sim1\tev$ higgsino region reproduces the measured Higgs boson
mass and the relic density, it lies in multi-TeV regions of both \mhalf\ and
\mzero, and will be completely out of reach of the LHC, although
within reach of dark matter searches in underground one tonne
detectors and in the Cherenkov
  Telescope Array (CTA)\cite{Roszkowski:2014wqa,Roszkowski:2014iqa}. Likewise, both in the AF
and the FP/HB regions, all superpartners will be for the most part too
heavy to be seen at the
LHC\cite{Baer:2012vr,Cohen:2013xda,Fowlie:2014awa,Roszkowski:2014wqa}.

The only cosmologically favored region that gives an acceptable Higgs
boson mass and remains partially accessible to the LHC is the \stauc
region.  In fact, numerous studies prior to the LHC,
\eg\cite{deAustri:2006pe}, found that the \stauc region was slightly
preferred over the other ones because lighter sleptons and
electroweakinos (EWinos) in loop corrections reduced the discrepancy
with the SM value of the anomalous magnetic moment of the
muon\cite{Bennett:2006fi}.  This motivates us to consider the
possibility that at \roots{14} the LHC might observe direct evidence
for a model in the \stauc region of the CMSSM.

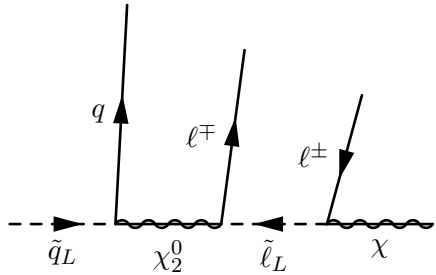
\begin{figure}
\centering
\input{GoldenDecay} 
\vspace{0.4cm} 
\caption{Feynman diagram for the golden decay.}
\label{Fig:GoldenDecay}
\end{figure}

In preparation for the LHC operation, various
groups\cite{Hinchliffe:1996iu,Miller:2005zp,Bachacou:1999zb},
including the ATLAS collaboration\cite{Aad:2009wy}, simulated the
precision with which the LHC could measure the masses of light
sparticles in the CMSSM. Sparticle masses can be measured from
kinematic endpoints in the invariant mass distributions of their decay
products\cite{Gjelsten:2004ki,Lester:705139} in the so-called ``golden
decay'' channel (\reffig{Fig:GoldenDecay}) which is a multi-stage
squark decay,
\begin{equation}\label{Eqn:DecayChain}
\s{q}_L \to \neut{2} q \to \s{\ell}^{\pm} \ell^\mp q \to \chi \ell^+ \ell^- q,
\end{equation}
where $\s{q}_L$ denotes a first- or second-generation left squark, $\chi^0_{1}=\chi$
and $\chi^0_{2}$ respectively the lightest and second-lightest neutralino and
$\slepton$ ($\ell$) a first- or second-generation slepton (lepton).
The experimental signature for this chain
will therefore be two leptons with opposite signs, a jet and missing
energy. The jet will usually be hard but it might not be if the mass
difference between the squark and the second-lightest neutralino happens to be
small. The chain results in striking kinematic endpoints in the
invariant mass distributions of the outgoing SM particles.

Because the masses of the leptons can be neglected, the position of the edge of
the leptons' invariant mass distribution is a function of only the relevant
sparticle masses;
\begin{equation}
m_{\ell\ell,\,\textrm{edge}}^2=
\left(
m_{\neut{2}}^2 - m_{\s{\ell}}^2
\right)
\left(
\frac{ m_{\s{\ell}}^2 - m_{\chi}^2 }{ m_{\s{\ell}}^2 }
\right).
\end{equation}
Measuring $m_{\ell\ell,\,\textrm{edge}}$ gives a relationship between
$m_{\neut{2}}$, $m_{\s{\ell}}$ and $m_{\chi}$. Measuring other
invariant mass edges allows a model-independent reconstruction of all the masses involved
in the decay chain\cite{Bachacou:1999zb}. The squark flavors,
however, will be indistinguishable, and a particular slepton might
dominate the decay chain.

Squark production and subsequent decay via the golden decay was
simulated in detail for particular CMSSM benchmark points, including
SPS1a\cite{Allanach:2002nj} and ATLAS SU3\cite{Aad:2009wy}, from which
the potential precision of sparticle mass measurements with a small
integrated luminosity of $1\invfb$ was
estimated\cite{Aad:2009wy,Gjelsten:2004ki,Lester:705139,Allanach:2001qe,Allanach:2003jw,Lafaye:2007vs,Bechtle:2009ty}.

In\cite{Roszkowski:2009ye}, the precision and reliability
with which Lagrangian parameters for the SU3 point in the CMSSM
could be determined were examined from a Bayesian perspective.  Assuming a simple Gaussian
approximation to the likelihood function describing golden decay
measurements (but using a proper covariance matrix), it was found that the
correct value of \mhalf was reproduced very well, while the correct values of
\mzero, \tanb and \azero were reproduced increasingly poorly, in that order. In the
case of the last variable, even the sign of \azero remained
undetermined. The effect of imposing in addition the neutralino's relic
density constraint was studied and found to be very important in
improving the reconstruction of \mzero and, to a lesser extent,
\tanb, while the error in \azero remained large. Sensitivity to the
choice of prior was also examined and found to be weak.

In an earlier version of this paper, the issue of parameter
reconstruction of the SU3 point was extended  using a similar methodology to models beyond the CMSSM:
the Non-Universal Higgs Model (NUHM), the CMSSM with non-universal
gaugino masses (CMSSM-NUG), both with two additional (although
different) free parameters, and finally the MSSM with twelve free
parameters, all defined at the EW scale. We propagated experimental
uncertainties in sparticle masses to uncertainties in SUSY Lagrangian
parameters with sophisticated statistical tools. We checked whether
the benchmark point was accurately recovered, or whether additional
information, from, for example, DM density would be required to
recover the Lagrangian parameters, and whether one could distinguish
various patterns of soft-breaking masses. Unsurprisingly, despite
assuming more realistic uncertainties in computing the neutralino's
relic density, in the CMSSM we basically reproduced the results of\cite{Roszkowski:2009ye},
while in the extended
models we found that the prospects of reconstruction were generally
poorer, although in some cases this was primarily so because the extended models relaxed gaugino mass
unification rather than because of their larger number of free parameters. In the extended
models prior dependence again became an issue. In a related paper\cite{Allanach:2011ya} the SU3 point with the CMSSM
mass spectrum was used to examine whether golden decay measurements at
the LHC could be accommodated instead by some other SUSY breaking
scenarios, with a generally positive conclusion.

Unfortunately, the SU3 point, along with most light SUSY scenarios,
was in fact excluded by direct searches for SUSY at the LHC at
\roots{7}\cite{Chatrchyan:2012uea,Aad:2012fqa}. Light SUSY scenarios are also
disfavored by the discovery of a Higgs boson with a mass $\mh \sim 126
\gev$\cite{Chatrchyan:2012ufa,Aad:2012tfa}, since heavy stops are
needed to produce large radiative corrections to the Higgs boson  mass; see,
\eg,\cite{Kowalska:2013hha}.

In this paper we return to anticipating that SUSY might be residing
not far above the current lower limits and that, after its hiatus, the LHC might discover SUSY in its
\roots{14} phase\cite{Baer:2012vr}. To this end, and in light of our earlier
discussion, we select a new benchmark point in the \stauc region
allowed by current LHC bounds, which we specify below. We ask, if nature is
described by a CMSSM scenario, compatible with previous direct
searches and with the Higgs boson observation, how well might one be able to measure
sparticle masses at the LHC and, subsequently, how well might one be
able to determine Lagrangian parameters and make predictions for other
observable quantities.

In light of LHC results the CMSSM might appear less
natural than before\cite{Fowlie:2014xha}; however, it correctly reproduces both the Higgs boson mass and
the DM relic abundance in the ``unnatural'' multi-TeV regions
of mass parameters. It also remains compatible with all experimental
data, with the exception of the anomalous magnetic moment of the muon.

Our paper is organized as follows; in
\refsec{sec:bm}, we construct our scenario, in \refsec{sec:exp}, we
describe the Monte-Carlo simulation of sparticle mass measurements via
the kinematic endpoints of a golden decay; in \refsec{sec:stat}, we
recapitulate our statistical methodology, with which we propagate
uncertainties in hypothetical sparticle mass measurements to
uncertainties in Lagrangian parameters, and in \refsec{sec:results}
we present our results.

\section{\label{sec:bm} Benchmark point and scenario}
First we identify a CMSSM parameter point that is compatible with existing
constraints and that might be found early in the LHC \roots{14} run
via the golden decay. In addition to the CMSSM parameters, there are SM nuisance
parameters with experimental uncertainties that can impact SUSY
phenomenology: the top pole mass, \mt, the bottom quark running mass,
\mbmbmsbar, and the strong and electromagnetic couplings, \alphas and
\invalpha respectively. Our complete set of parameters is, therefore,
\begin{equation}\label{eqn:params}
\params = \left(\mhalf, \mzero, \azero, \tanb, \sgnmu, \mt,
\mbmbmsbar,\alphas, \invalpha\right).
\end{equation}

Our desiderata are that our benchmark point
exhibits a golden decay, predicts a relic density in agreement with
Planck\cite{Ade:2013xsa} and a correct Higgs boson mass, $\mh \sim 126
\gev$, within theory errors, and is consistent with constraints from
current direct SUSY searches.

For a squark to cascade via the golden decay a particular
sparticle mass hierarchy is required, and for the process to be significant  necessitates
other (spoiler) decay modes to be suppressed and kinematic phase space
to be moderate\cite{Gjelsten:2004ki}. We thus require that the gluino
is heavier than the squarks, shutting the $\s{q} \to q \s{g}$ spoiler
mode. We obviously require that squarks are heavier than the second
lightest neutralino, which is typical in the CMSSM. We require that
the second lightest neutralino is heavier than a slepton; although the
second lightest neutralino could reach an identical final state via a
three-body decay mediated by a $Z$-boson, the kinematic endpoints
would be ruined. In fact, we require that it is at least $50\gev$
heavier, so that the decay is not suppressed by small phase-space.

Because we wish to simultaneously agree with DM density
measurements\cite{Ade:2013xsa}, our benchmark must invoke a particular
mechanism to annihilate DM in the early Universe. The aforementioned
requirement, that $\miii > \mii + 50 \gev$, limits the CMSSM parameter
space from which we can select our benchmark to regions in which
$\mhalf\gg\mzero$. This requirement in conjunction with that from the
relic density, means we can pick only from the \stauc region of CMSSM
parameter space, in which neutralinos and staus are almost degenerate
in mass and efficiently coannihilate. We, however, require that
the neutralino and stau masses differ by more than the $\tau$-lepton mass to avoid limits from
long-lived staus\cite{Citron:2012fg}. Within the \stauc region, current
direct SUSY search constraints impose $\mhalf \gsim 850
\gev$\cite{ATLAS-CONF-2013-047,Aad:2014wea}.

\begin{figure}[ht]
\centering
\includegraphics[width=0.8\linewidth,keepaspectratio=true]{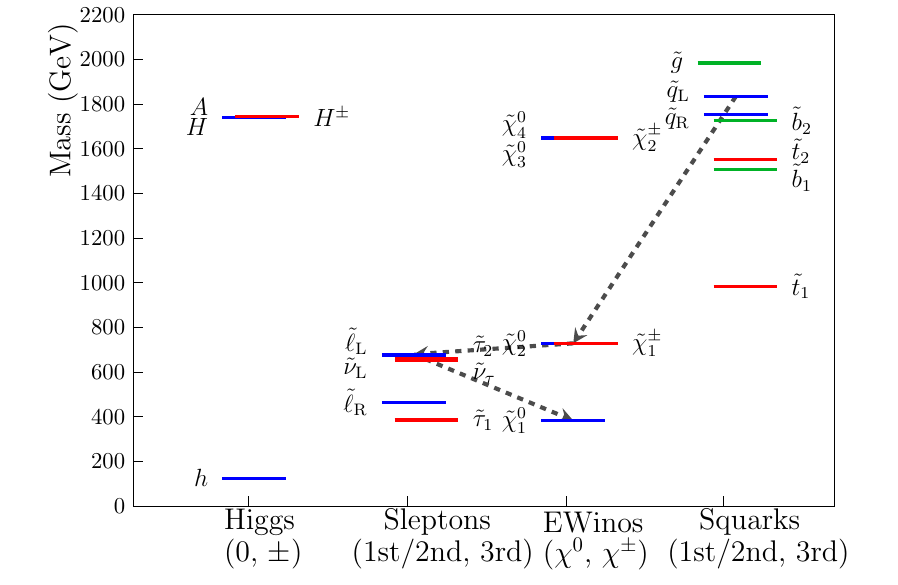}
\caption{\label{fig:spec}The mass spectrum for our benchmark point,
    with the golden decay indicated by arrows, calculated with
    \texttt{softsusy-3.3.7}\cite{Allanach:2001kg}. Because the first-
    and second-generation sparticles are approximately degenerate in
    mass, they are plotted only once each.}
\end{figure}

We located our benchmark point by conducting a local scan within a
small part of the parameter space in the \stauc region with the \texttt{Minuit}
algorithm\cite{James:1975dr}. We fixed $\mhalf=900\gev$ to maximize
production cross sections but still stay above the lightest permitted
value in\cite{ATLAS-CONF-2013-047,Aad:2014wea}, and varied the CMSSM's three
other continuous parameters and the SM top pole mass to minimize a
$\chi^2$-function from nine experiments, including Higgs boson, dark matter
and $B$-physics experiments. This way we found a local minimum which
we accepted as our benchmark point. The point and its mass spectrum
are shown in \reftable{tab:bm} and \reftable{tab:spec}, respectively,
and its mass spectrum is plotted in \reffig{fig:spec}. Our benchmark's
branching ratios are such that a golden decay begins most often with a
left-handed first- or second-generation squark, and proceeds via a
left-handed first- or second-generation slepton. At the electroweak
scale, our benchmark has $\mu\sim1.5\tev \gg M_1, M_2$. The lightest
and second-lightest neutralinos are approximately entirely bino- and wino-like,
respectively. Our benchmark top pole mass $\mt = 174.3\gev$ is larger than but
in agreement with its world average in\cite{Beringer:1900zz}, especially if one
follows caveats regarding the interpretation of top mass measurements as
measurements of the pole mass.

\begin{table}[ht]
\centering
\begin{tabular}{|c|c|c|}
\toprule
Parameter & Description & Benchmark value \\
\colrule
\mzero   & Unified scalar mass      & $315$\gev\\
\mhalf   & Unified gaugino mass     & $900$\gev\\
\azero   & Unified trilinear        & $-2550$\gev\\
\tanb    & Ratio of Higgs vevs      & $11.0$\\
\sgnmu   & Sign of Higgs parameter  & $+1$\\
\mt      & Top pole mass            & $174.3$\gev\\
\mbmbmsbar& Bottom running mass     & $4.18$\gev\\
\invalpha& Inverse of EM coupling   &  $127.944$\\
\alphas  & Strong coupling          & $0.1184$\\
\botrule
\end{tabular}
\caption{Our benchmark CMSSM parameters with three significant figures
    and benchmark SM nuisance parameters. The SM nuisance
    parameters are the world averages in\cite{Beringer:1900zz}, with the
    exception of \mt. }
\label{tab:bm}
\end{table}

\begin{table}[ht]
\begin{center}
\begin{tabular}{|p{1.25cm}p{1.25cm}|p{1.25cm}p{1.25cm}|p{1.25cm}p{1.25cm}|p{1.25cm}p{1.25cm}|}
\toprule
\multicolumn{8}{|c|}{Particle Mass (GeV):}\\
\colrule
$\neut{1}=\chi$& $382.8$   & $\s{e}_L$       & $679.8$ & $\s{d}_L$ & $1835$ & $h$ & $124.1$\\
\neut{2}       & $728.7$   & $\s{e}_R$       & $463.4$ & $\s{d}_R$ & $1754$ & $H$ & $1741$\\
\neut{3}       & $1645$    & $\s{\nu}_e$     & $675.1$ & $\s{u}_L$ & $1834$ & $A$ & $1742$\\
\neut{4}       & $1649$    & $\s{\tau}_1$    & $384.6$ & $\s{u}_R$ & $1762$ & $H^\pm$ & $1744$\\
\charg{1}      & $728.9$   & $\s{\tau}_2$    & $659.9$ & $\s{b}_1$ &$1509$  & &\\
\charg{2}      & $1649$    & $\s{\nu}_\tau$  & $651.4$ & $\s{b}_2$ & $1726$ & &\\
\s{g}          & $1985$    &                 &         & $\s{t}_1$ & $984.1$& &\\
               &           &                 &         & $\s{t}_2$ & $1552$ & &\\
\botrule
\end{tabular}
\caption{\label{tab:spec}The particle mass spectrum for our CMSSM
    benchmark, calculated with
    \texttt{softsusy-3.3.7}\cite{Allanach:2001kg}. The first- and
    second-generation sparticles are approximately degenerate in mass.}
\end{center}
\end{table}

Since our benchmark point lies in the \stauc region, $m_{\s{\tau}_1}
\approx \mchi$ (\reftable{tab:spec}), though their masses are chosen
to differ by more than the $\tau$-lepton mass to avoid limits from
long-lived staus\cite{Citron:2012fg}. With this restriction, and
$\mhalf \gsim 850 \gev$ from direct searches, the relic density cannot
be reduced via \stauc to the Planck measurement, $\abund=0.1196$; we
can go as far down as $\abund=0.1390$\cite{Belanger:2010gh}.

We check, however, that our
benchmark point is in global agreement with experimental
constraints. In an ensemble of measurements, there is an appreciable
chance of a discrepant measurement. Our benchmark point satisfied nine
experimental measurements: the Higgs boson mass, dark matter relic density, \bsgamma,
    \butaunu, \bsmumu, \mw, \sinsqeff, \mt and \delmbs.,\footnote{
      Our benchmark Higgs boson mass is somewhat on a low side.
Note, however, that is has been computed at the two-loop
    level. Recently computed three-loop level corrections due to a
    resummation of leading and sub-leading logarithms in the top/stop
    sector\cite{Hahn:2013ria} increase the value by roughly 1\gev for
    mass spectra of the order of our benchmark point.}
having fitted four parameters, with
$\chi^2=8.8$. Assuming that our benchmark point describes nature, the
probability of obtaining by chance $\chi^2\ge8.8$, the $p$-value, is
$12\%$. Our benchmark point is therefore acceptable.\footnote{Our benchmark point cannot
explain a small anomaly in a CMS search\cite{CMS:2014jfa,Allanach:2014gsa}, which observed
a dilepton edge at about $80\gev$, because our benchmark point predicts about $215\gev$.}

\section{\label{sec:exp}Simulating golden decay measurements}
One can consider the Minkowski-square of sums of the four-momenta of
the visible SM decay products in \refeq{Eqn:DecayChain}, resulting in
four invariant masses:
\begin{align}
\label{Eqn:Momenta}
m_{\ell\ell}^2 &= (p_{\ell_\text{near}} + p_{\ell_\text{far}})^2,\\\nonumber
m_{\ell_\text{near}q}^2 &= (p_q + p_{\ell_\text{near}})^2,\\\nonumber
m_{\ell_\text{far}q}^2 &= (p_q + p_{\ell_\text{far}})^2,\\\nonumber
m_{\ell\ell q}^2 &= (p_q + p_{\ell_\text{near}} + p_{\ell_\text{far}})^2,
\end{align}
where $\ell_\text{near}$ and $\ell_\text{far}$ are the first and
second lepton produced in a cascade, respectively.
Because at the LHC one cannot identify the lepton's origins, we instead
consider the maximum and minimum invariant mass from the quark-lepton
combinations. Four kinematic endpoints are predicted by calculating
these Minkowski-squares and maximizing with respect to angular
distributions. The clearest endpoint is that from $m_{\ell\ell}^2$,
because it has no spin corrections as it is mediated by a scalar.
The locations of these four kinematic features can be exactly
predicted from the sparticle mass spectrum; the formulas were derived
in\cite{Lester:705139,Gjelsten:2004ki}.

We simulated 10,000 SUSY events at $\roots{14}$ with our
benchmark point with the \texttt{Pythia 8.1}\cite{Sjostrand:2007gs}
Monte-Carlo event generator. This number of events is equivalent to
$\sim100\invfb$, which could be collected in $\sim2$ years at the
LHC. We applied geometrical cuts to insure events were inside the
detector, but otherwise applied no detector simulation. We selected
events with an opposite-sign same-flavor lepton pair, at least two
jets (with standard definitions for leptons and
jets) and missing energy. We required that the hardest jet had transverse momentum $p_T>100\gev$ and
removed the $Z$-boson peak in the leptonic invariant mass distribution by
vetoing $89\gev< m_{\ell\ell} <95\gev$.

In the CMSSM, because of $R$-parity conservation, colored sparticles ought to be produced in pairs in QCD interactions at the LHC; each side of a SUSY QCD event
should contain a hard jet. We had to decide which of the two hardest jets was associated with the golden decay. If we were looking for a high (low) edge in an invariant mass distribution, we chose the jet that minimized (maximized) that
invariant mass. This choice was conservative; invariant mass
distributions were never populated beyond their expected edges because
of contamination with the jet from the opposite side of the decay chain.

We produced histograms for our selected events according to the invariant masses
in \refeq{Eqn:Momenta} with \texttt{ROOT}\cite{Brun:1997pa}. The
resulting invariant mass distributions exhibited the expected
kinematic features. With \texttt{Minuit}, we extracted their positions
by fitting simple curves to the distributions with a least-squares
technique, with Poisson $\sqrt{N}$ error bars in our histograms. This
resulted in uncorrelated pseudo-measurements of four endpoints with
statistical errors. Because the statistical error is dominant, we
neglected systematic errors in our analysis.

To find sparticle masses from the endpoints, we used formulas for the
endpoints in\cite{Lester:705139,Gjelsten:2004ki}. The measurable masses are:
\begin{equation}\label{eqn:atlasset}
\mchi, m_{\neut{2}}, m_{\s{q}}, m_{\s{\ell}},
\end{equation}
where $m_{\s{q}}$ and $m_{\s{\ell}}$ are effective squark and slepton masses that
dominate the golden decay. In our earlier SU3 studies, the effective squark mass was the
average of the first- and second-generation squark masses, because at the LHC their
productive cross sections are substantial and one cannot easily distinguish
first- and second-generation squarks, and the effective slepton mass
was that of the lightest slepton, which was assumed to dominate
the decay chain;
\begin{eqnarray}
m_{\s{q}}  &=& \frac{1}{8}\left(
    m_{\tilde{u}_R} + m_{\tilde{u}_L} +
    m_{\tilde{d}_R} + m_{\tilde{d}_L} +
    m_{\tilde{s}_R} + m_{\tilde{s}_L} +
    m_{\tilde{c}_R} + m_{\tilde{c}_L}
\right),\\
m_{\s{\ell}}  &=& \min \left(
    m_{\tilde{e}_R}, m_{\tilde{e}_L},
    m_{\tilde{\mu}_R}, m_{\tilde{\mu}_L},
    m_{\tilde{\tau}_1}, m_{\tilde{\tau}_2}
\right).
\end{eqnarray}

For our current study we refined our definitions. The effective squark mass was
the average of only left-handed first- and second-generation squarks, and the
effective slepton mass was the average left-handed selectron and smuon masses;
\begin{eqnarray}
m_{\s{q}}  &=& \frac14\left(
    m_{\tilde{u}_L} +
    m_{\tilde{d}_L} +
    m_{\tilde{s}_L} +
    m_{\tilde{c}_L}
\right),\\
m_{\s{\ell}}  &=& \frac12 \left(
    m_{\tilde{e}_L}+
    m_{\tilde{\mu}_L}
\right).
\end{eqnarray}
Because the second lightest neutralino is wino-like for our benchmark point, the
golden decay is dominated by left-handed sparticles. Because taus cannot be reliably
detected, they were omitted from our analysis. The definitions are model dependent; in relaxed
models, the golden decay might be dominated by other sparticles.

Rather than using
the inverted formulas for sparticle masses, however, we fitted
sparticle masses to our pseudo-measurements of the endpoints by
minimizing a $\chi^2$-function with \texttt{Minuit}. The
\texttt{Migrad} minimization algorithm in \texttt{Minuit} is a
quasi-Newton method, which utilizes derivatives of the
$\chi^2$-function. The matrix of second derivatives of the
$\chi^2$-function is proportional to the inverse covariance
matrix. The resulting covariance matrix, defined by ${\sigma}_{ij} =
{\sigma}_{ji} = \text{E}(X_i)\text{E}(X_j) - \text{E}(X_i X_j)$,
where $\text{E}$ denotes an expectation value and
\begin{equation}
X=(\mi, \mii, \miii,\miv)^T,
\end{equation}
is
\begin{equation}
\sigma =
\begin{pmatrix}
132.0 & 18.4 & 31.9 & 175.8\\
\cdot   & 25.5 & 24.2  & 21.3\\
\cdot   & \cdot   & 24.8 & 39.6\\
\cdot   & \cdot   & \cdot   & 401.1 \\
\end{pmatrix},
\end{equation}
in units of $(\text{GeV})^2$.
The non-zero off-diagonal covariance
matrix elements indicate that the mass measurements are correlated. The $\chi^2$-function associated with the covariance matrix is
\begin{equation}
\chi^2 = (X - X_\text{BM})^T \sigma^{-1} (X - X_\text{BM})
\end{equation}
where $X_\text{BM}$ stands for $X$ evaluated at our benchmark point, \ie, the expected values of $X$.

For insight, we diagonalize the inverse of this matrix to obtain the
orthonormal eigenvectors, the combinations of masses that can be
independently measured, and their eigenvalues. For clarity, let us make our original basis,  $(\hmi,\hmii,\hmiii\,\hmiv)$,
explicit;
\begin{equation}
X=\mi\cdot\hmi+\mii\cdot\hmii+\miii\cdot\hmiii+\miv\cdot\hmiv.
\end{equation}
To diagonalize the inverse covariance matrix, we find the orthogonal matrix $V$ such that $V^T\sigma^{-1}V$ is diagonal. The errors for the
independent mass combinations are the eigenvalues of $\sigma^{-1}$, \ie, the  entries of $V^T\sigma^{-1}V$;
\begin{equation}
V\,\sigma^{-1}\, V^T \approx \text{diag}\[(0.3 \gev)^{-2},(5.6 \gev)^{-2},(7.5 \gev)^{-2},(22.3 \gev)^{-2}\],
\end{equation}
and the independent combinations of masses are the eigenvectors of $\sigma^{-1}$, \ie, the columns of the orthogonal matrix that diagonalizes our covariance matrix;
\begin{align}
\label{Eqn:Vecs}
V_{1i} &= 0.1 \cdot \hmi +0.7 \cdot \hmii -0.8 \cdot \hmiii + 0.0 \cdot \hmiv,\\\nonumber
V_{2i} &= 0.6 \cdot \hmi -0.6 \cdot \hmii -0.4 \cdot \hmiii -0.2 \cdot \hmiv,\\\nonumber
V_{3i} &= 0.6 \cdot \hmi -0.4 \cdot \hmii -0.5 \cdot \hmiii +0.4 \cdot \hmiv,\\\nonumber
V_{4i} &= 0.4 \cdot \hmi +0.1 \cdot \hmii +0.1 \cdot \hmiii +0.9 \cdot \hmiv.
\end{align}
The combinations of masses that can be independently measured are:
\begin{align}
V_{1i} X_i &= 0.1 \cdot \mi +0.7 \cdot \mii -0.8 \cdot \miii + 0.0 \cdot \miv
\approx \frac{1}{\sqrt{2}}(\mii - \miii),\\\nonumber
V_{2i} X_i &= 0.6 \cdot \mi -0.6 \cdot \mii -0.4 \cdot \miii -0.2 \cdot \miv,\\\nonumber
V_{3i} X_i &= 0.6 \cdot \mi -0.4 \cdot \mii -0.5 \cdot \miii +0.4 \cdot \miv,\\\nonumber
V_{4i} X_i &= 0.4 \cdot \mi +0.1 \cdot \mii +0.1 \cdot \miii +0.9 \cdot \miv
\approx \miv.
\end{align}
It will later be of significance that the combination $(\mii -
\miii)$ has by far the smallest experimental uncertainty. We assume
that the expected measured endpoints are our benchmark's endpoints
(\ie, that there is no bias), and that consequently, the expected
extracted masses are our benchmark's sparticle masses.

\section{\label{sec:stat}Statistical methodology}
We first recapitulate our Bayesian methodology; more details can be
found in, \eg,\cite{Roszkowski:2006mi,Fowlie:2011mb}. At each point
$\params=(\mzero,\mhalf,\azero,\tanb)$ in the CMSSM's parameter space
we calculate our observables.  We want to know the posterior
probability density function (pdf) of the CMSSM parameters
given the experimental data $\datalr$. We find this via Bayes'
theorem,
\begin{align}
\label{Eqn:BayesTheorem}
p(\params|\datalr) &= \frac{p(\datalr|\params)\, p(\params)}{p({\datalr})}\nonumber,\\
&=\frac{\likef{\params} p(\params)}{\ev},
\end{align}
where \likef{\params} is the likelihood function, the probability
of obtaining data $\datalr$ for our observables at a given point $\params$;
\priorf{\params} is the prior probability density function, our prior
belief in the CMSSM parameter space and nuisance parameters; \ev is
the Bayesian evidence, which is just a normalization factor in this instance; and
$p(\params|d)$ is the posterior pdf, the object that we wish to find,
given the experimental data $\datalr$.

We will obtain the posterior with the nested sampling
algorithm\cite{Skilling06}, implemented in the \texttt{MultiNest}
computer package\cite{Feroz:2007kg,Feroz:2008xx}, by supplying the
algorithm with our chosen priors for the CMSSM parameters and with our
likelihood functions for the experimental data. We will inspect the
posterior by plotting $68\%$ and $95\%$ credible regions on
two-dimensional planes of the CMSSM parameter space. We find such
regions by first marginalizing the posterior over the other two
CMSSM parameters and over all SM nuisance parameters,
\begin{equation}\label{Eqn:Marginalise}
\probg{x_1,x_2}{\datalr} = \int \probg{\params}{\datalr}\,\text{d}x_3\text{d}x_4\ldots,
\end{equation}
and second finding the smallest regions of the $\(x_1, x_2\)$ plane
that contain $68\%$ and $95\%$ of the marginalised posterior; these
are the credible regions,
\begin{equation}
\int \probg{x_1,x_2}{\datalr} \text{d}x_1\text{d}x_2 = 0.68 \text{~or~} 0.95.
\end{equation}

We will consider three sets of experimental data:
\begin{enumerate}
\item hypothetical golden decay sparticle mass measurements described in \refsec{sec:exp} (\lhc);
\item Higgs boson mass data (\higgs);
\item dark matter relic density (\planck).
\end{enumerate}

Our likelihood function for \lhc, described in \refsec{sec:exp}, is a
multivariate Gaussian, reflecting the correlations in the sparticle
mass measurements,
\begin{equation}\label{Eqn:LikeLHC}
\likef{\lhc} = \exp\[-0.5\, (X - X_\text{BM})^T \sigma^{-1} (X - X_\text{BM}) \],
\end{equation}
where ${\sigma}^{-1}$ is the inverse of the covariance matrix (note
that this is already a squared quantity) obtained in \refsec{sec:exp},
and  $X$, already defined above, is the set of the measurable sparticles masses evaluated at the
CMSSM point in question, whilst the index BM denotes that the quantity
is evaluated at our benchmark point.

We calculate the sparticle mass spectrum and the Higgs boson mass $\mh$ with
\texttt{softsusy 3-3.7}\cite{Allanach:2001kg}, and the relic density $\abundchi$
of the lightest neutralino with
\texttt{micromegas-2.4.5}\cite{Belanger:2010gh}.  These quantities are
compared with measurements through likelihood
functions, which are assumed to be Gaussian, though we include in
quadrature theoretical errors in the CMSSM predictions for these
observables of $3\gev$\cite{Allanach:2004rh} and
$10\%$\cite{Allanach:2004jh,Allanach:2004jy}, respectively:
\begin{align}
\likef{\higgs} &= \gaussian{125.8\gev}{\mh}{[(0.6\gev)^2 + (3\gev)^2]}\nonumber,\\
\likef{\planck} &= \gaussian{0.1196}{\abundchi}{[0.0031^2+(0.1\abundchi)^2]},
\end{align}
where the means and standard deviations are those reported by
CMS\cite{Chatrchyan:2012ufa,Aad:2012tfa} and by
Planck\cite{Ade:2013xsa}, respectively.  The relic density of
$\abundchi=0.1390$ and Higgs boson mass of $124.1\gev$ calculated at our
benchmark point agree, within theoretical and experimental errors,
with the Planck measurement\cite{Ade:2013xsa} $\abund=0.1196$ and from
the CMS measurement\cite{Chatrchyan:2012ufa,Aad:2012tfa} of
$125.8\gev$, respectively.
We, however, include in our likelihoods the
measured relic density and measured Higgs boson mass, rather than our
benchmark's values. This is not a fault; it reflects the experimental
and theoretical uncertainties in these measurements and any biases
that they might introduce.

We consider three cases, by adding these data one by one:
\begin{enumerate}
\item golden decay pseudo-data only, $\like = \likef{\lhc}$;
\item golden decay and Higgs boson mass data, $\like = \likef{\lhc} \cdot \likef{\higgs}$;
\item golden decay, Higgs boson mass and  relic density data, $\like =
  \likef{\lhc} \cdot \likef{\higgs} \cdot \likef{\planck}$.
\end{enumerate}

Our choice of priors for the CMSSM parameters is slightly moot; our
likelihood \likef{\lhc} ought to be informative enough to overcome
different, sensible choices of prior\cite{Trotta:2008bp}. We choose
flat priors for the CMSSM parameters that evenly weight the parameter
space, listed in \reftable{tab:priors}, but expect our posterior to be
independent of this choice.

In general we should also vary the SM nuisance parameters but their
inclusion would be computationally expensive and would have limited
impact in our scenario. We fix these parameters to the weighted
averages of their experimental values in\cite{Beringer:1900zz}, rather than our benchmark point's values.
Our benchmark point's top mass in
\reftable{tab:bm}, however, is slightly larger than the weighted average to maximise the Higgs boson mass.
In our scenario, one would know only the measured top mass and not our benchmark point's top mass.

\begin{table}[ht]
\begin{center}
\begin{tabular}{|c|c|c|}
\toprule
Parameter & Prior range & Distribution\\
\colrule
\mzero          & $(0.1,\, 4)\tev$  & Flat\\
\mhalf          & $(0.1,\, 2)\tev$  & Flat\\
\azero          & $(-4,\, 4)\tev$   & Flat\\
\tanb           & $(3,\, 62)$       & Flat\\
\sgnmu          & $+1$              & Fixed\\
\mt             & $173.5$\gev       & Fixed\\
\mbmbmsbar      & $4.18$\gev        & Fixed\\
\invalpha       & $127.944$         & Fixed\\
\alphas         & $0.1184$          & Fixed\\
\botrule
\end{tabular}
\caption{\label{tab:priors} Our prior distributions for the CMSSM
  parameters in our calculation of the posterior pdf for
  the CMSSM. The SM nuisance parameters are fixed to their
  world averages in\cite{Beringer:1900zz}.}
\end{center}
\end{table}

\section{\label{sec:results}Results}

We successfully found the posterior pdf and subsequently the credible regions for the CMSSM in three cases: our hypothetical \lhc likelihood only, our hypothetical \lhc likelihood and a likelihood from the \higgs mass measurement, and our hypothetical \lhc likelihood and likelihoods  from the \higgs mass measurement and the \planck relic density measurement. To highlight the impact and necessity of additional information in parameter reconstruction, we discuss our results parameter plane by parameter plane in all three cases, rather than likelihood by likelihood.

\begin{figure}[ht]
\centering
\subfloat[][\lhc likelihood only]{\label{fig:m0m12:i}
\includegraphics[width=0.32\linewidth, clip=True, trim=0 0 0 2cm]{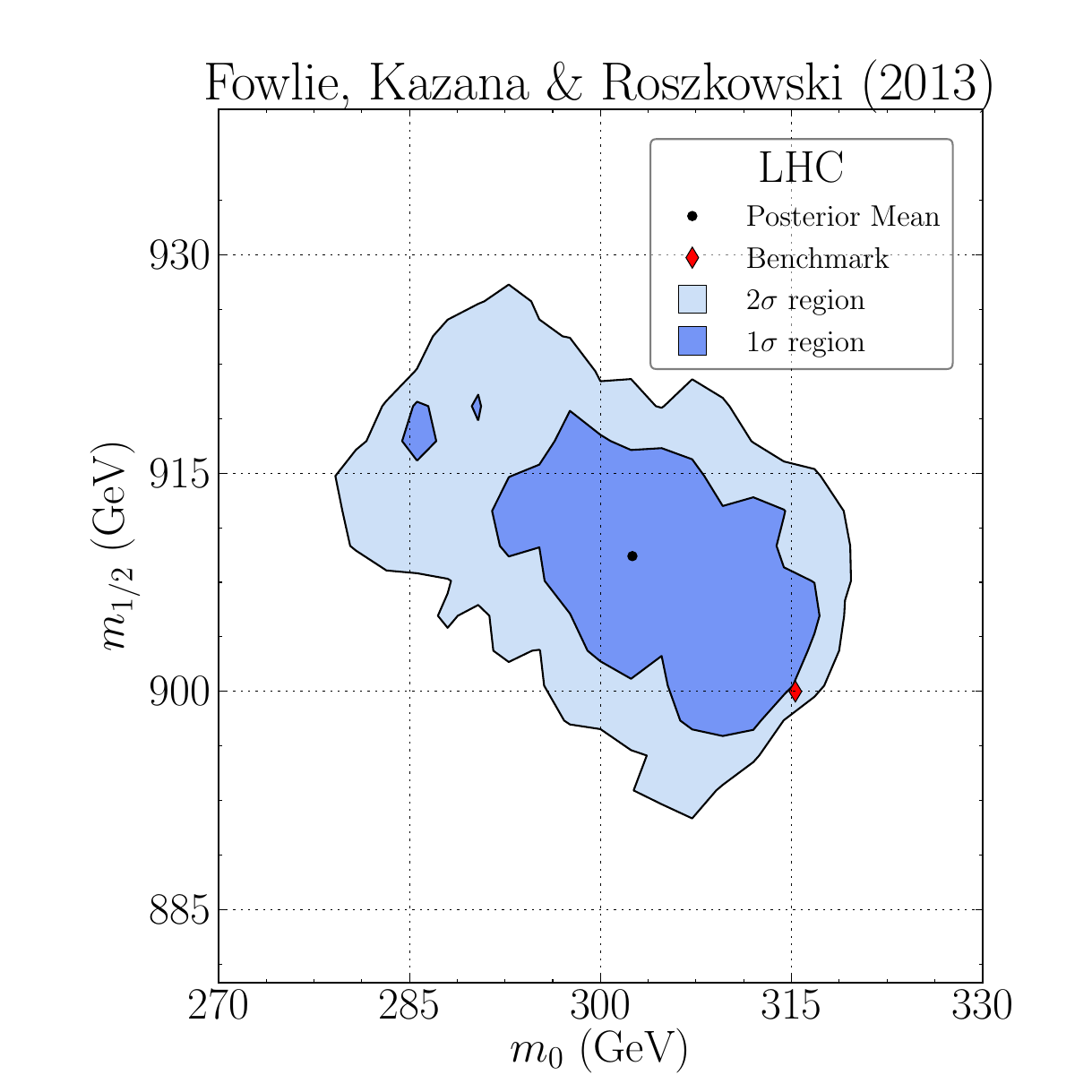}
}
\subfloat[][\lhc+\higgs likelihoods]{\label{fig:m0m12:ii}
\includegraphics[width=0.32\linewidth, clip=True, trim=0 0 0 2cm]{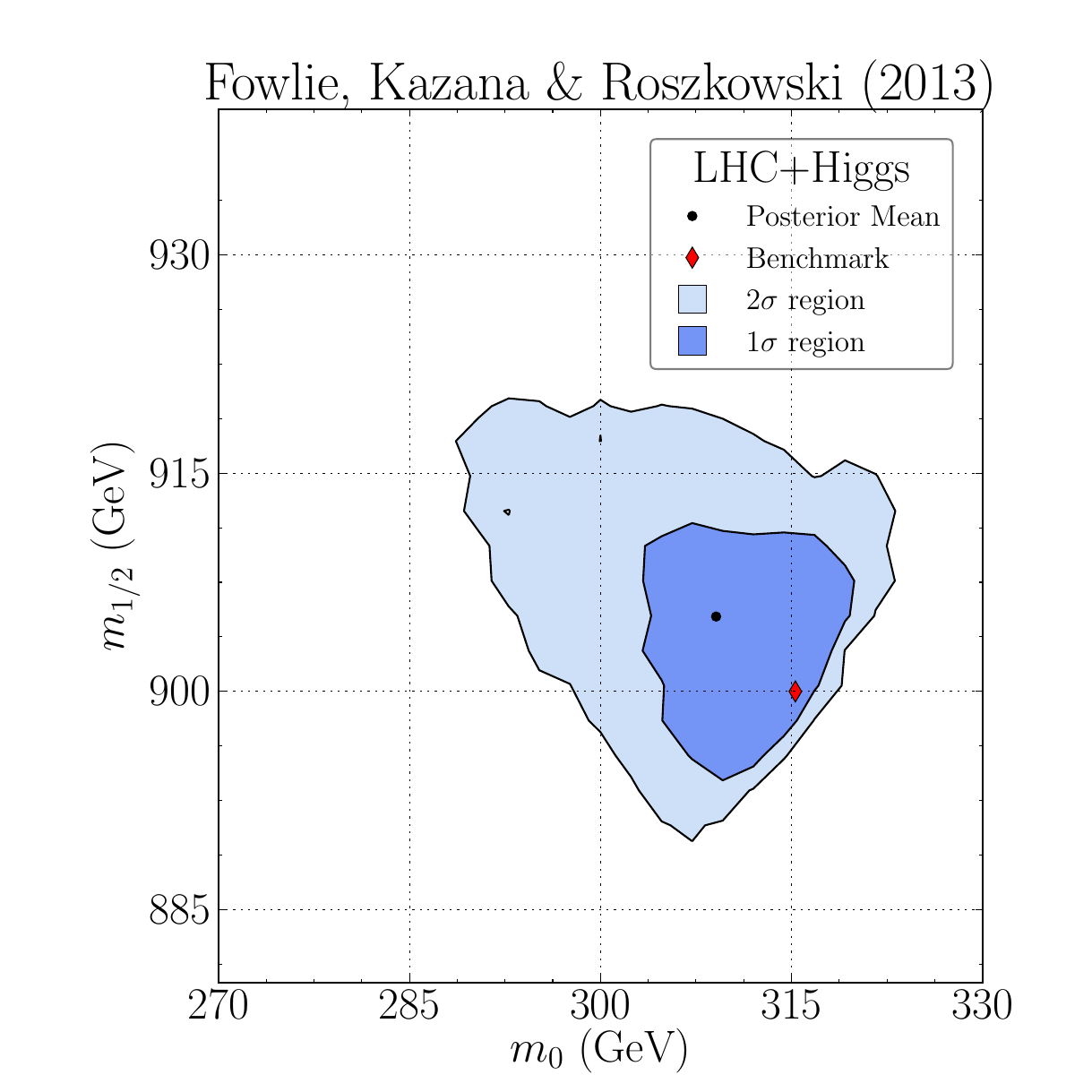}
}
\subfloat[][\lhc+\higgs+\planck likelihoods]{\label{fig:m0m12:iii}
\includegraphics[width=0.32\linewidth, clip=True, trim=0 0 0 2cm]{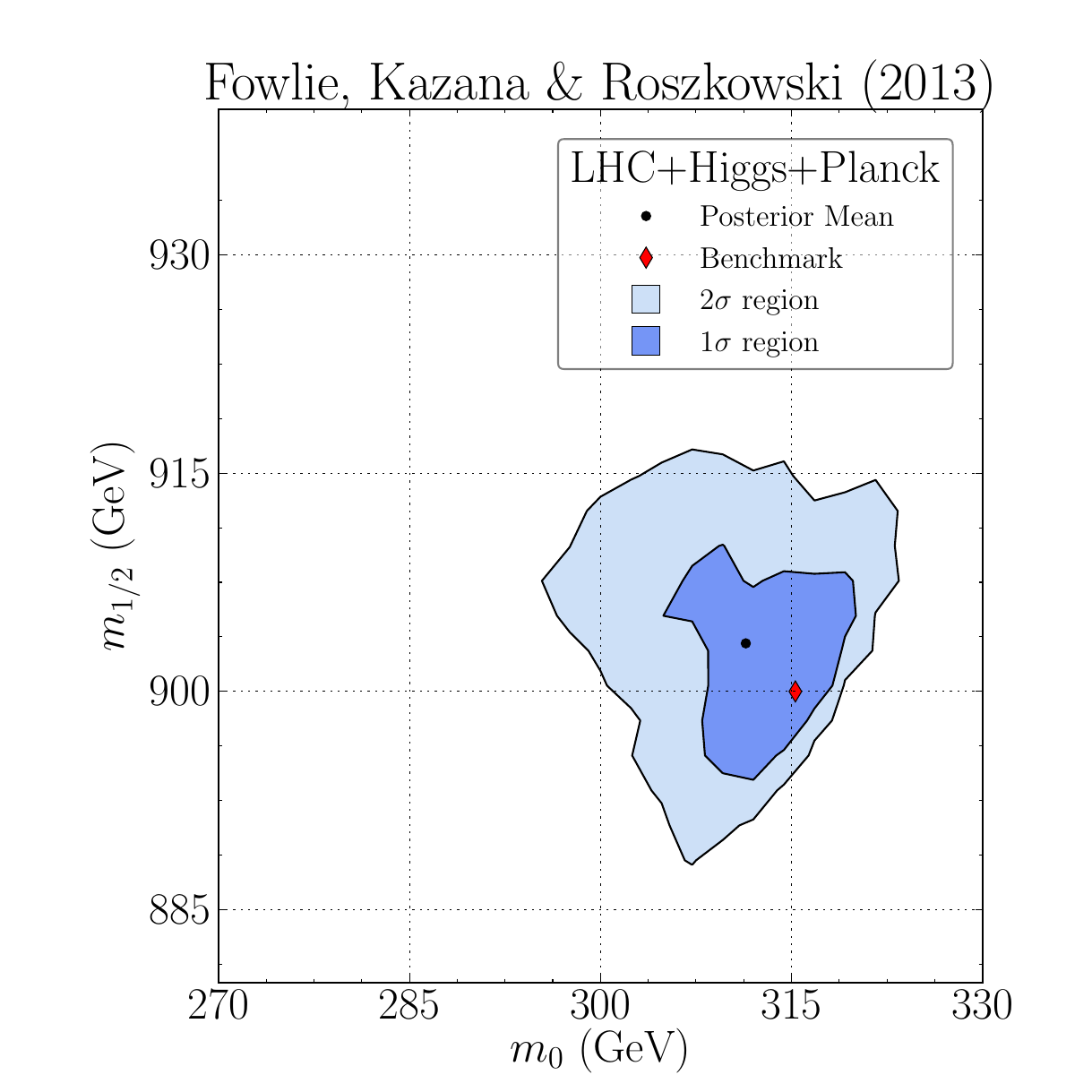}
}
\caption{The $68\%$ (dark blue) and $95\%$ (light blue) credible regions of the \pmm plane of the CMSSM, adding the data one by one from left to right. The parameters \tanb and \azero were marginalised. Our benchmark point is marked with a red diamond and the posterior mean is marked with a black circle. The bin limits are identical to the plot limits with $25$ bins per dimension.}
\label{fig:m0m12}
\end{figure}

\begin{figure}[ht]
\centering
\subfloat[][\lhc likelihood only]{\label{fig:a0tanb:i}
\includegraphics[width=0.32\linewidth, clip=True, trim=0 0 0 2cm]{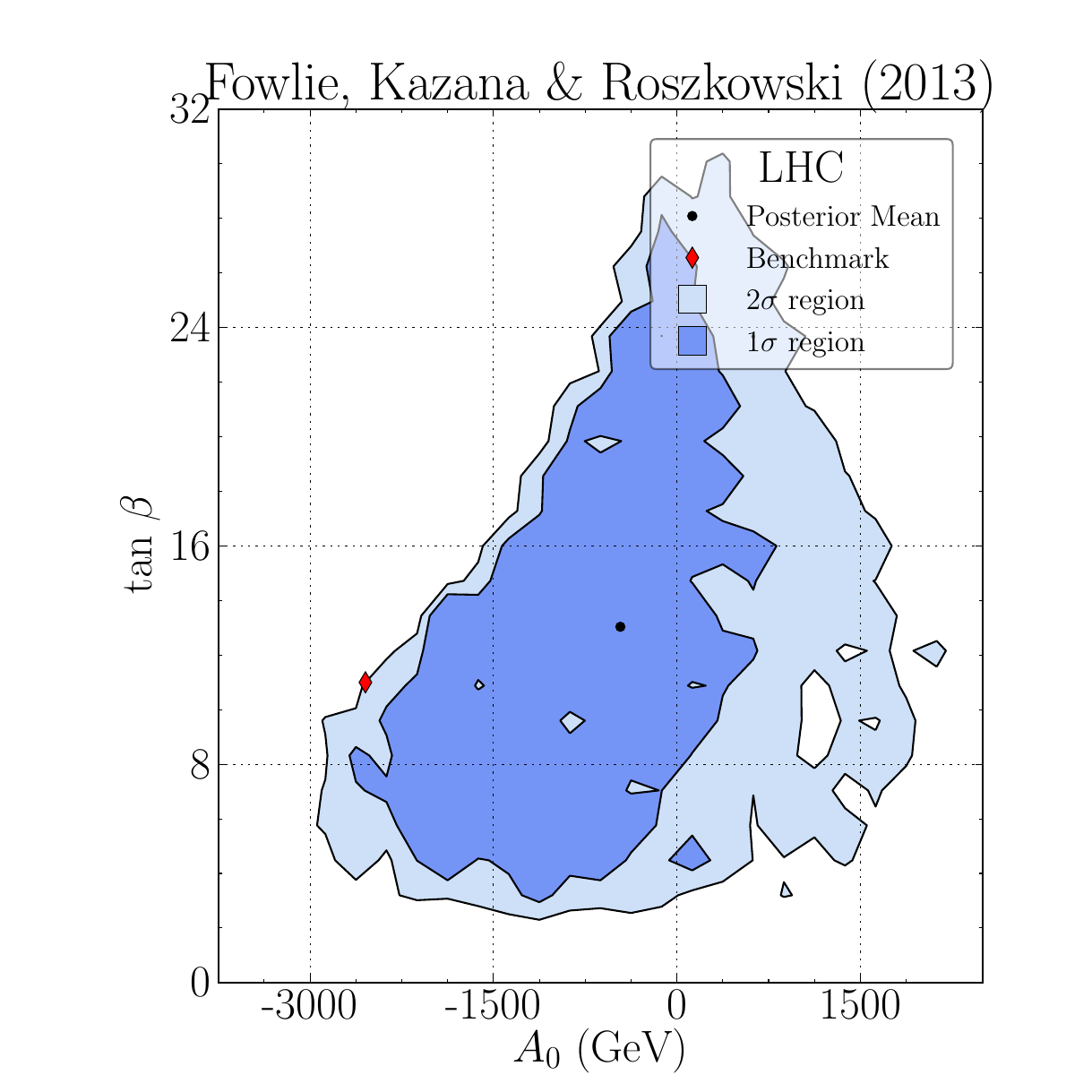}
}
\subfloat[][\lhc+\higgs likelihoods]{\label{fig:a0tanb:ii}
\includegraphics[width=0.32\linewidth, clip=True, trim=0 0 0 2cm]{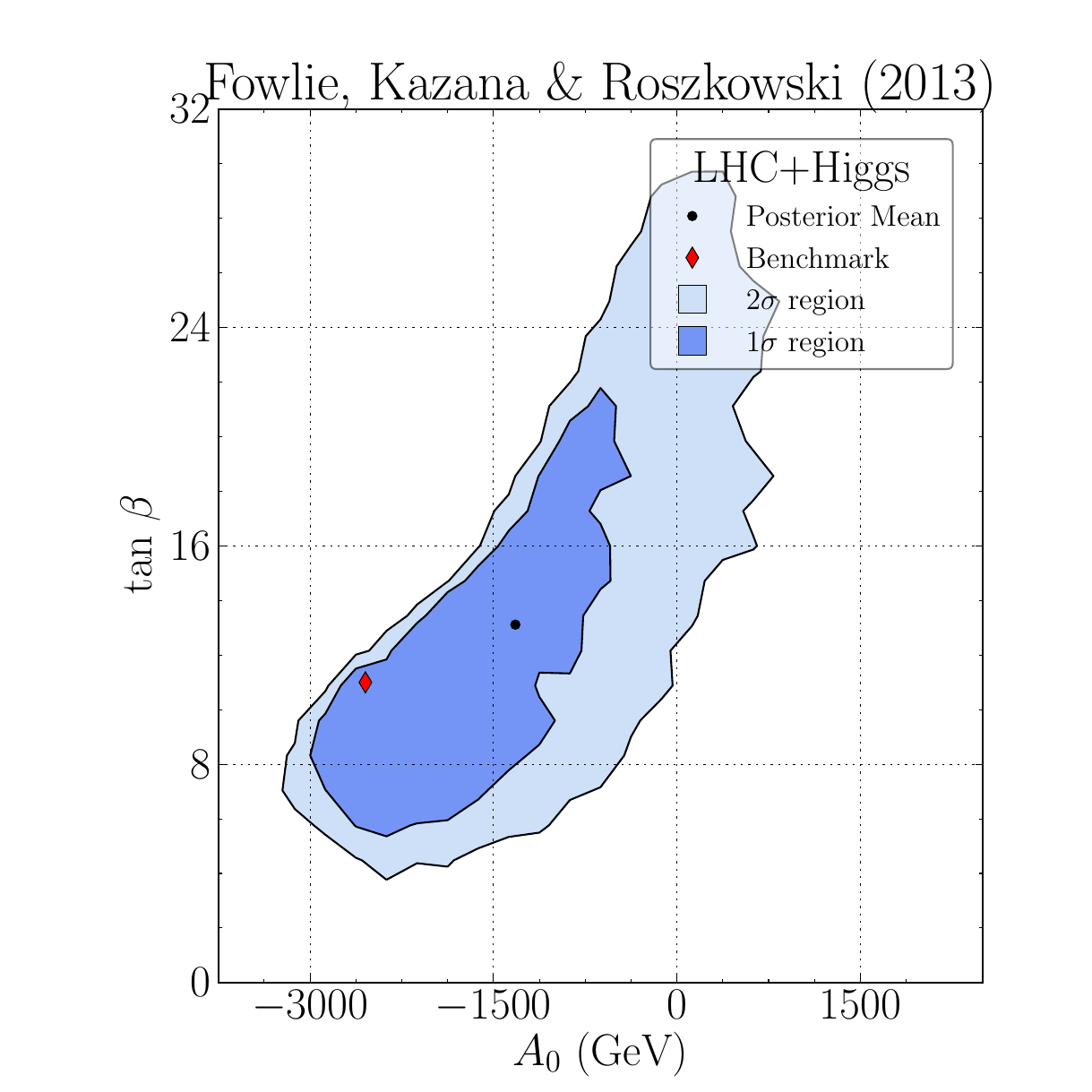}
}
\subfloat[][\lhc+\higgs+\planck likelihoods]{\label{fig:a0tanb:iii}
\includegraphics[width=0.32\linewidth, clip=True, trim=0 0 0 2cm]{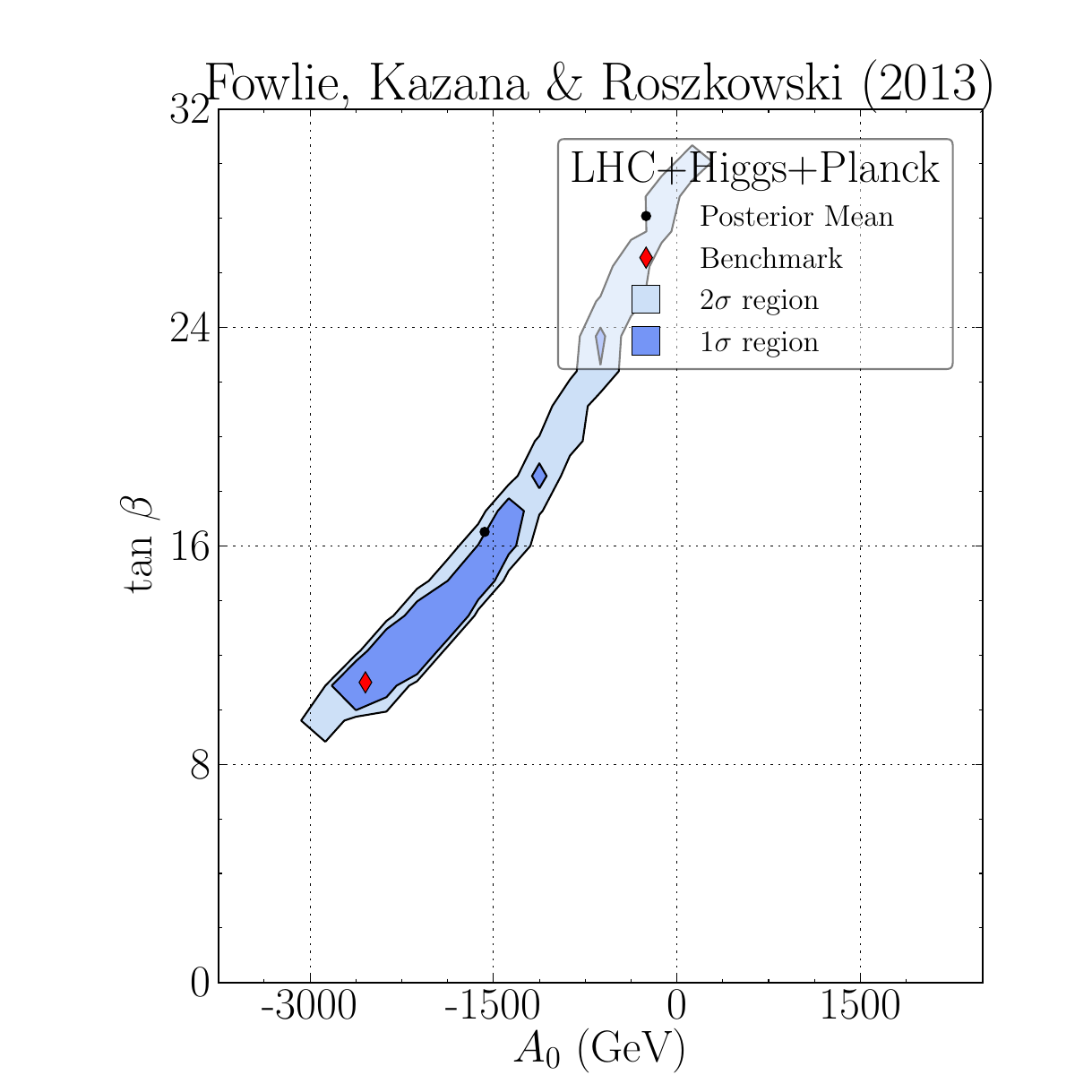}
}
\caption{The $68\%$ (dark blue) and $95\%$ (light blue) credible regions of the \pat  plane of the CMSSM, adding the data one by one from left to right. The parameters \mhalf and \mzero were marginalized. Our benchmark point is marked with a red diamond and the posterior mean is marked with a black circle. The bin limits are identical to the plot limits with $25$ bins per dimension.}
\label{fig:a0tanb}
\end{figure}

We first consider the \pmm plane of the CMSSM in \reffig{fig:m0m12},
showing all three cases from left to right. (The corresponding
results in the \pat plane of the CMSSM are shown below in \reffig{fig:a0tanb}.)
Note the narrow $60\gev
\times 60\gev$ scales shown; the parameters are reconstructed at the
$95\%$ level to within $\lesssim 20\%$. Reconstruction of \pmm with
only \lhc is somewhat successful; a single $95\%$ mode is recovered,
which envelopes the benchmark point. Our benchmark point is, however,
outside our $68\%$ credible region. Two orthogonal directions in the
parameter space are visible: the anti-diagonal $\mzero-\mhalf$
direction and the diagonal $\mzero+\mhalf$ direction, which correspond
to the first and second eigenvectors of our covariance matrix in
\refeq{Eqn:Vecs}.

To see that the $\mzero-\mhalf$ direction corresponds to the first
eigenvector, recall that the first eigenvector corresponded to the
$\mii - \miii$ combination of masses with a small $0.3\gev$
uncertainty.  That combination is approximately unchanged if \mzero
and \mhalf are simultaneously increased by $\sim10\gev$, as each mass
is increased by a similar amount, since $\mii \approx 0.8\mzero$.  This dominant measurement,
therefore, constrains our credible regions to a line
$\mhalf\propto\mzero$, with a breadth proportional to the $0.3\gev$
uncertainty.  The $\mzero-\mhalf$ direction, perpendicular to this
line, is constrained.

To see that the $\mzero+\mhalf$ direction corresponds to the second
eigenvector, recognise there is no approximate cancellation in $V_{2i}X_i$ if
\mhalf and \mzero are simultaneously increased, but that cancellations
are possible if \mhalf is increased and \mzero is decreased. This
measurement, therefore, constrains our credible regions to a line
$\mhalf\propto-\mzero$, with a breadth proportional to the $5.6\gev$
uncertainty. The $\mzero+\mhalf$ direction, perpendicular to this
line, is constrained.

The third and fourth eigenvectors are negligible; the third because that combination of masses changes relatively slowly with \pmm, because of cancellations in $\mi-\miii$ and $\mii-\miv$, and the fourth because its associated uncertainty is large.

Because of its small associated uncertainty of $0.3\gev$, we might
expect that the anti-diagonal $\mzero-\mhalf$ direction is more
constrained than the diagonal $\mzero+\mhalf$ direction on the \pmm plane. The second-lightest
neutralino mass, however, can be tuned independently of the slepton
mass by $\sim10\gev$ by tuning $\azero$ to increase the
second-lightest neutralino's higgsino component. Note that in
\reffig{fig:a0tanb:i} $\azero$ is not well constrained.
This freedom broadens
the credible region in the $\mzero-\mhalf$ direction. One can decrease
\mzero from its benchmark and increase \mhalf from its benchmark, to
keep \mii constant, but achieve $\miii-\mii$ similar to its
pseudo-measurement by increasing $\azero$ from its benchmark which
decreases \miii to compensate for the increase in \mhalf. One cannot
compensate for increased \mzero and decreased \mhalf by decreasing
$\azero$, however, because \neut{2} mass is already saturated at our
benchmark, \ie~it is already entirely wino-like. This is why the
$\mzero+\mhalf$ direction is more constrained.

The credible regions shrink successively as the data is added, though
two orthogonal directions in the parameter space remain
visible. The $\mzero+\mhalf$ direction of the credible region is only
marginally shrunk by additional data, whereas the $\mzero-\mhalf$
direction of the credible region is squashed. When we add \higgs,
\azero must be $\lesssim0.5\tev$ (see \reffig{fig:a0tanb:ii}) to
increase the Higgs boson mass via maximal mixing. The left-hand side
of the \pmm credible region, which achieved agreement with \lhc via
$\azero\sim0$, is excluded and absent in
\reffig{fig:m0m12:ii}. Increases in Higgs boson mass from increasing
\mhalf and \mzero to increase stop masses are negligible. Indeed,
varying \mhalf by $50\gev$ ($10\gev$) of our benchmark point changes
the Higgs boson mass by only $0.1\gev$ ($0.0\gev$), which is
negligible compared with the $3\gev$ theory error in \mh.

When we add \planck, we enforce $m_{\s{\tau}_1} \approx \mi$, so that
staus and neutralinos coannihilate effectively and reduce the relic
density to the \planck value. This further squashes the
$\mzero-\mhalf$ direction of the credible region. Furthermore, \planck
requires that \pat are tuned so that the stau mixing results in
$m_{\s{\tau}_1} \approx \mi$, reducing the freedom to tune the
neutralino masses with \azero for the \lhc pseudo-measurements (compare \reffig{fig:a0tanb:iii}). This
is rather fortunate; \higgs and \planck constrain the direction of
parameter space that was poorly constrained by \lhc, enhancing the
impact of this additional information.

The complicated dependence of neutralino and stau masses on \azero
results in bias in our credible regions and posterior means. In
\reffig{fig:m0m12}, in the $\mzero-\mhalf$ direction, our credible
regions are not centered on our benchmark point. This asymmetry
ultimately results from the asymmetry in $\azero$; whilst decreases
from its benchmark increase the second-lightest neutralino's higgsino
component, increases are ineffectual, because the second-lightest
neutralino is already entirely wino-like. If our statistics were
unbiased, the posterior mean would equal the benchmark point in all
cases, and the credible regions would shrink around the benchmark as
data was added. However, we approach the correct benchmark as data is
added (our statistics are consistent).

The \pat plane of the CMSSM in \reffig{fig:a0tanb} tells a different
story from that of the \pmm plane. With a likelihood from \lhc in
\reffig{fig:a0tanb:i}, \pat are poorly reconstructed; our credible
regions exhibit a single, broad mode that omits the benchmark
point. The unskewed shape of the credible regions indicate little
correlation between \tanb and \azero, though a slight positive
correlation is visible. The reconstruction is significantly worse than
for \pmm, especially for \azero, which is determined to within
$-3.5\tev\lesssim\azero\lesssim1.5\tev$ at the $95\%$ level.

This is unsurprising. Because our \lhc likelihood constrains only
neutralino and first- and second-generation sparticle masses, at
tree-level, it is independent of \azero and dependent on \tanb only in
off-diagonal neutralino mass matrix elements from $F$-terms and
diagonal sfermion mass matrix elements from $D$-terms. To first order,
our credible regions on the \pat plane with only \lhc are determined
by physicality conditions. Large $\tanb\gtrsim30$ is excluded, because
the stau is lighter than the neutralino. Similarly, large negative $\azero$ is disfavored at large \tanb, because in that corner of the \pat plane off-diagonal trilinear and $F$-terms in sfermion mass matrices do not cancel.

Our credible regions, however, favor $\azero\lesssim0$, similar to its
benchmark, whereas physicality ought to marginally favor
$\azero\gtrsim0$. The preference for $\azero\lesssim0$ stems from
$\mu$ in the neutralino mass matrix. The wino-like, second-lightest
neutralino's mass is
\begin{equation}
\label{Eq:M2}
\miii \approx M_2 -\frac{M_W^2\(M_2+\mu\sin2\beta\)}{\mu^2-M_2^2},
\end{equation}
where $\mu$ is determined from the electroweak symmetry breaking condition.
If $\mu$ decreases, the second-lightest neutralino's higgsino
component increases, and its mass decreases. The soft-breaking Higgs
boson masses receive one-loop corrections proportional to the
trilinear soft terms corresponding to the Yukawa couplings for the top and bottom quarks
$|A_{t,b}|^2$, respectively, in their renormalization group equations from squark
loop diagrams. Ultimately, \azero affects $\mu$ via the
renormalization group and an electroweak symmetry breaking condition, which affects neutralino
masses. Thus, our \lhc likelihood is somewhat sensitive to \azero.

Incidentally, there is no positive \azero solution. Our benchmark has $A^\text{BM}_{t,b}\lesssim-2\tev$ at the electroweak scale. No equivalent solution with $|A^\text{BM}_{t,b}|$ exists. The $\beta$-functions for $A_{t,b}$ are negative and run quicker than that for $A_{\tau}$. For $|A^\text{BM}_{t,b}|$ at the electroweak scale, we would require $\azero\gtrsim10\tev$. But such large \azero is vetoed, because the stau is lighter than the neutralino.

The \pat plane with only \lhc requires $\azero\gtrsim-3\tev$ at the
$95\%$ level. With larger negative $\azero$, the stau is the LSP, or, even if one simultaneously decreases $\tanb\lesssim6$ to avoid stau LSP, the neutralino mass is already saturated at our benchmark. The correction in \refeq{Eq:M2} is always negative. With our benchmark \azero, the correction is already approximately zero, because the neutralino is predominantly wino-like. With larger negative $\azero$, we cannot fine-tune the neutralino mass with \azero. In other words, because the second-lightest neutralino is entirely wino-like at our benchmark, one cannot decrease its higgsino component. With $\azero\gtrsim0.5\tev$, with different corrections to bino- and wino-like neutralino masses, similar to the second term in \refeq{Eq:M2}, it becomes impossible to tune \pmm to maintain agreement with the pseudo-measurements.

Adding information from \higgs (\reffig{fig:a0tanb:ii}) helps somewhat
improve reconstruction on the \pat plane, with \azero pushed down to
negative values and \tanb pushed to slightly higher values, without exceeding $\tanb
\lesssim 30$ at $95\%$ established by \lhc. The absolute value of
\azero is pushed heavier and \tanb is pushed higher to tune
stop-mixing so that $X_t / \msusy \approx -\sqrt{6}$,\footnote{At our
  benchmark, $X_t / \msusy \approx -2.1$.} which maximizes the Higgs
boson mass, bringing it closer to its measured value.

Adding information from \planck (\reffig{fig:a0tanb:iii}) dramatically
improves parameter reconstruction on the \pat plane, though reveals
strong positive correlation between \pat. The relic density is most
sensitive in this region of the CMSSM to the stau mass, in contrast to
our \lhc likelihood, which is sensitive to only the first- and
second-generation sleptons. Because the $\tau$ Yukawa coupling is significantly
larger than the $e$ and $\mu$ Yukawa couplings, the stau is sensitive to
mixing between left- and right-handed states, which splits the stau
mass eigenvalues. This mixing is proportional to $X_\tau \equiv
A_{\tau}-\mu\tanb$, and hence $m_{\s{\tau}_1}$ can be driven smaller
so that is approximately mass degenerate with \mi by increasing \tanb
or by increasing $|A_{\tau}|$. This enhances \stauc, thus decreasing
the relic density to its measured \planck value. Because we require
particular $X_\tau = \hat X_\tau \pm \Delta X_\tau$ to tune the
lightest stau's mass,
\begin{align}
\hat X_\tau - \Delta X_\tau &\lesssim  A_{\tau}-\mu\tanb \lesssim \hat X_\tau + \Delta X_\tau,\\\nonumber
A_{\tau}- \hat X_\tau - \Delta X_\tau &\lesssim \mu\tanb  \lesssim A_{\tau} - \hat X_\tau + \Delta X_\tau,
\end{align}
\tanb must lie between close parallel lines on the \pat plane, truncated at $\azero\sim0$ and $\azero\sim-3\tev$ by limits established by our \lhc and \higgs likelihoods in \reffig{fig:a0tanb:ii}.

Similarly to our statistics for \pmm, our statistics for \azero are biased, in that our distributions are not centered on our benchmark \azero. The cause is identical to that for \pmm: decreases in \azero from its benchmark result cannot increase the neutralino mass, because it is already saturated at our benchmark.

\begin{figure}[ht]
\centering
\subfloat[][\lhc likelihood only]{\label{fig:chisig:i}
\includegraphics[width=0.32\linewidth, clip=True, trim=0 0 0 2cm]{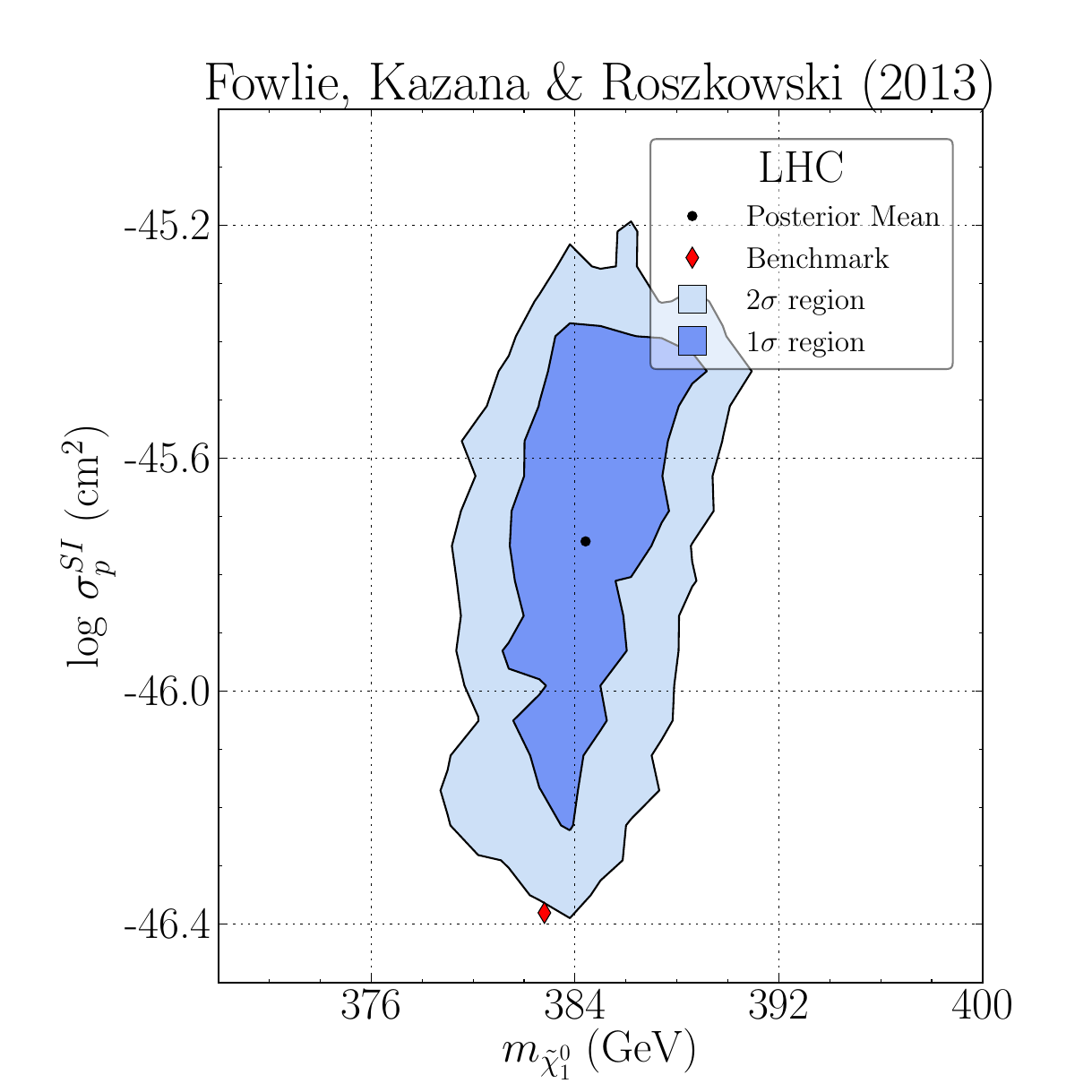}
}
\subfloat[][\lhc+\higgs likelihoods]{\label{fig:chisig:ii}
\includegraphics[width=0.32\linewidth, clip=True, trim=0 0 0 2cm]{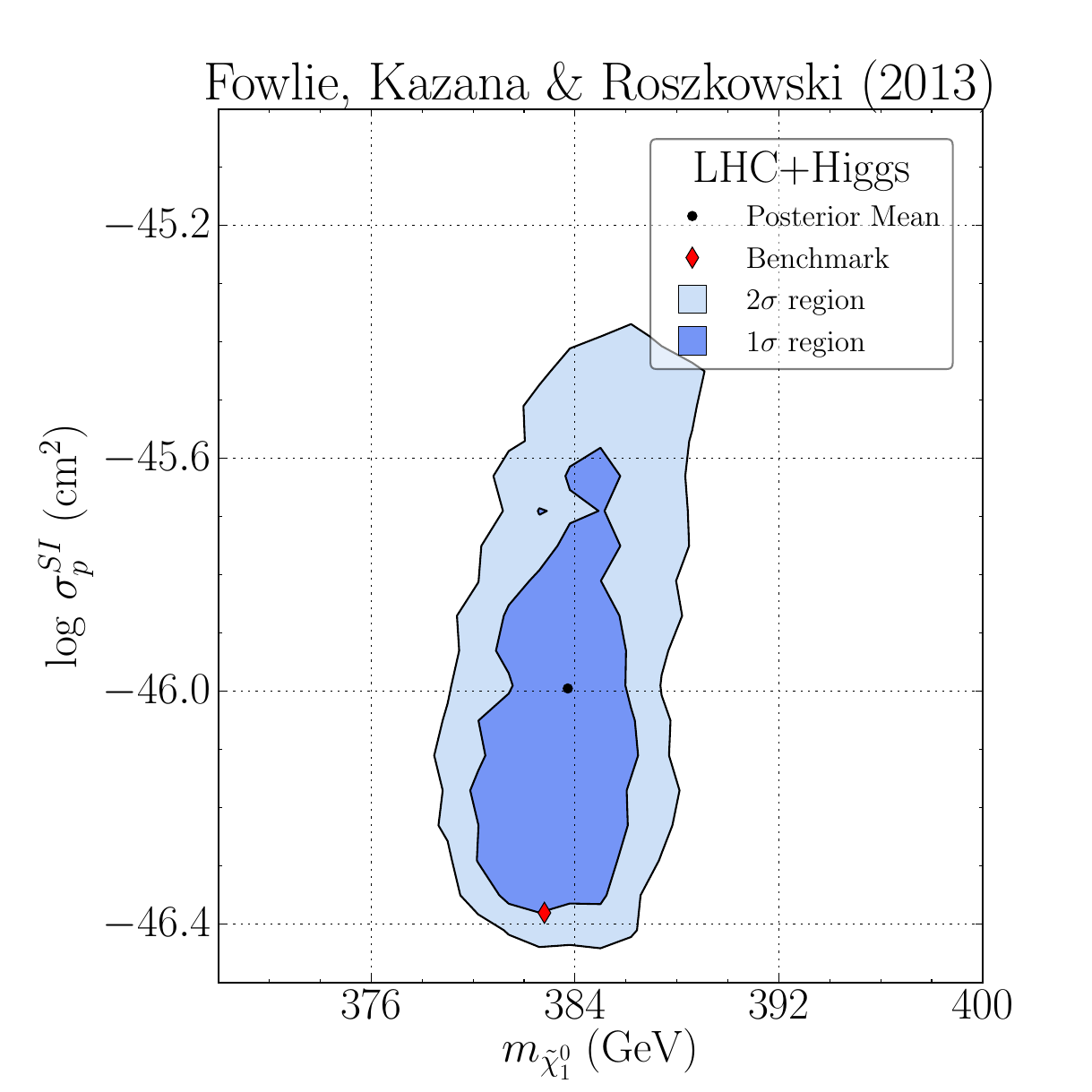}
}
\subfloat[][\lhc+\higgs+\planck likelihoods]{\label{fig:chisig:iii}
\includegraphics[width=0.32\linewidth, clip=True, trim=0 0 0 2cm]{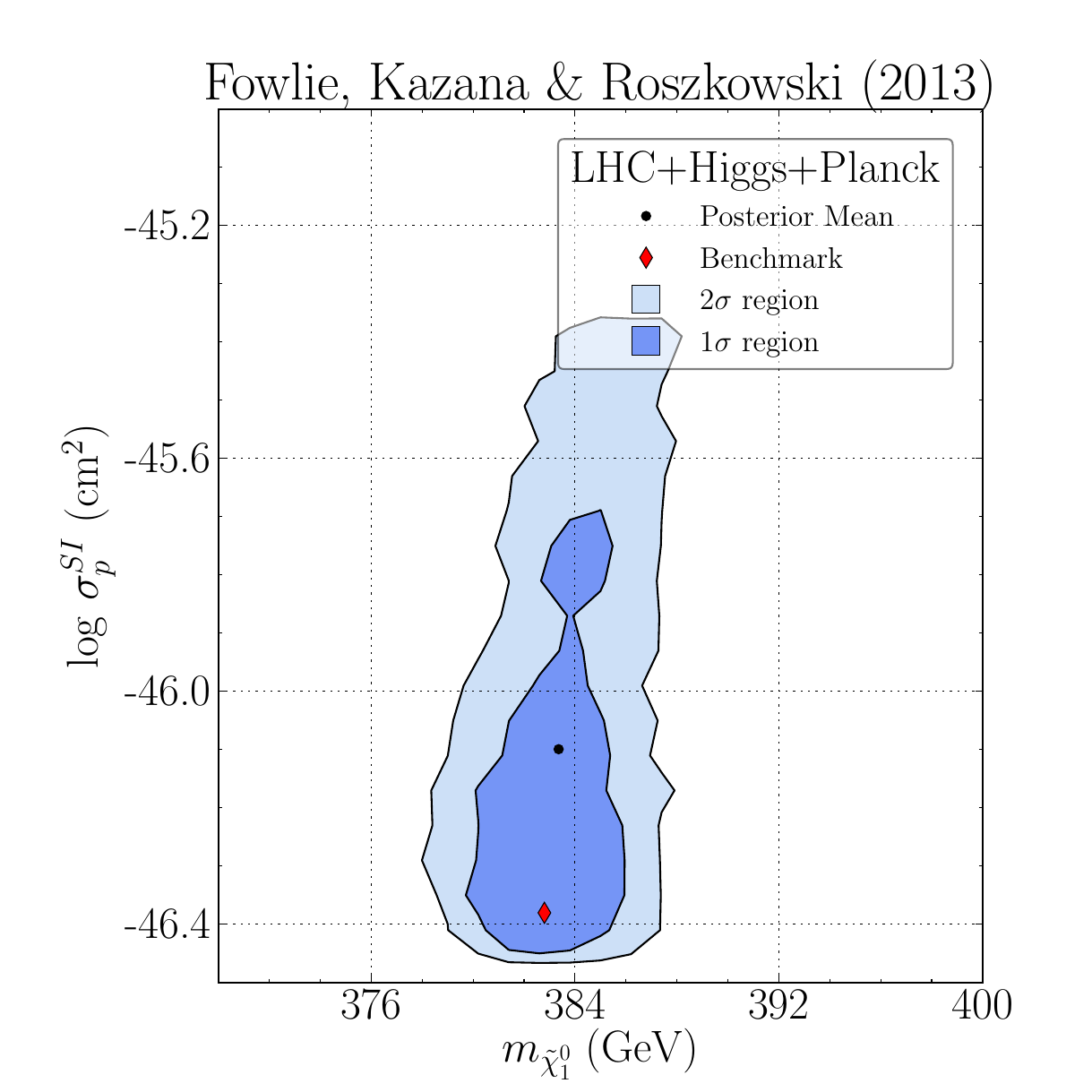}
}
\caption{The $68\%$ (dark blue) and $95\%$ (light blue) credible regions of the \pcs  plane of the CMSSM, adding the data one by one from left to right. The posterior for these derived quantities was obtained by histogramming our samples\cite{deAustri:2006pe}. Our benchmark point is marked with a red diamond and the posterior mean is marked with a black circle. The bin limits are identical to the plot limits with $25$ bins per dimension. The LUX limit (not shown) is $\sigsip\lesssim10^{-45}\,\text{cm}^2$.}
\label{fig:chisig}
\end{figure}

We turn to the experimental quantities that would be of much interest
in our scenario in which SUSY had been discovered. First, we consider
DM direct detection experiments, which would attempt to verify SUSY as
the DM. We plot credible regions on the \pcs plane in
\reffig{fig:chisig}. We found the posterior for these derived
quantities by histogramming our samples weighted by the
posterior\cite{deAustri:2006pe}. The plot shows little correlation
between \mi and \sigsip, and \mi and \sigsip are both reasonably
well-determined. Although \sigsip is determined at the $95\%$ level to
within an order of magnitude, with only \lhc, its distribution is biased towards
\sigsip larger than its benchmark. The increasing \sigsip direction
corresponds to the $\mhalf-\mzero$ direction on the \pmm plane. As
previously explained, the down-type higgsino component of the lightest
neutralino increases along that direction on the \pmm plane and thus
\sigsip increases.

The resolution and bias of \sigsip improves slightly as data is added,
especially \planck, but the resolution of \mi is not much improved by
the additional information. Nevertheless, the precision of the CMSSM
direct detection predictions indicate that in our discovery scenario
we would know that DM might be within reach of direct detection
experiments in the foreseeable future and be able to decide which
experiments to build accordingly. Our scenario might be inaccessible
at LUX and Xenon100, but should be accessible at a 1-tonne detectors
whose reach is expected to be below $10^{-46}\,\text{cm}^2$
in the neutralino mass range typical for our benchmark point. On the
other hand, prospects for current or future gamma-ray experiments,
like CTA, look hopeless\cite{Roszkowski:2014wqa,Roszkowski:2014iqa}.

Finally, we consider the rare decay \brbsmumu, which, if it deviates
from its SM value, could indicate the presence of new physics.
LHCb\cite{Aaij:2013aka} and CMS\cite{Chatrchyan:2013bka}, recently measured \brbsmumu with statistical
significance but limited precision as $\brbsmumu = 3.1\pm0.7
\times 10^{-9}$\cite{Beringer:1900zz}. We found that in our scenario one could make a
CMSSM prediction for \brbsmumu that agreed with the SM prediction within a $\sim 10\%$
uncertainty, stemming from uncertainties in the sparticle mass
spectrum and parametric uncertainties in SM nuisance parameters. Unfortunately, it probably
would not provide a channel for independently verifying our benchmark point.

\section{Conclusions}\label{Sec:Conclusions}
Assuming that the CMSSM is a viable model, we identified a benchmark point in the CMSSM's
heavy \stauc region that is in agreement with experimental constraints. We
demonstrated that our benchmark could be discovered at the LHC at
\roots{14} with $300\invfb$ via the golden decay, in which a squark
decays via a slepton to opposite-sign-same-flavor leptons, a jet and
missing energy. We simulated invariant mass distributions from this
decay in Monte-Carlo. From kinematic endpoints in these distributions,
we showed that we could measure the masses of two lightest neutralinos, a
squark and a slepton, with a methodology identical to that in
preliminary LHC studies. We found that these measurements were quite
precise, with small, correlated errors, described by our covariance
matrix.

We investigated, with Bayesian statistics, whether in our benchmark
scenario we could determine the CMSSM's Lagrangian parameters from the
sparticle mass measurements. CMSSM parameters can be accurately determined; our Bayesian credible regions enveloped the
benchmark parameters, though \azero was recovered with limited
precision, and the credible regions indicated bias. To investigate the
impact of Higgs boson mass and relic density measurements, we added
likelihoods describing these measurements, and repeated our Bayesian
analysis. We found that the additional experiments had limited impact,
though helped to determine \azero. Lastly, we considered whether, in
our benchmark scenario, we could make precise predictions for
experimental observables, with which one could verify that the
kinematic endpoints were from the CMSSM, and that the CMSSM's
neutralino was dark matter. We found that the spin-independent proton
scattering cross section, relevant to the direct detection of dark
matter, could be predicted to within an order of magnitude and
will be accessible to oncoming one tonne detectors. A prediction for the
branching ratio for the rare decay \bsmumu, however, with heavier,
approximately decoupled sparticles and small \tanb, was limited by
parametric errors from SM nuisance parameters and errors in the
CMSSM's Lagrangian parameters.


\begin{acknowledgements}
  A.J.F.~is supported by grant TK120 from the ERDF CoE program. L.R.~is supported by
  the Welcome Programme of the Foundation for Polish Science and in
  part by an STFC consortium grant of Lancaster, Manchester and
  Sheffield Universities.
\end{acknowledgements}


\bibliography{fkr1_vDIS}

\end{document}

%% file: fkr1macros.tex
\newcommand{\newc}{\newcommand*}

\long\def\begincomment#1\endcomment{%
        \begingroup\sf\baselineskip12pt#1\endgroup}

\newc{\etal}{\textrm{et al.}} 
\newc{\eg}{\textrm{e.g.}} 
\newc{\ie}{\textrm{i.e.}}
\newc{\etc}{\textrm{etc}}
\newc\vs{\textrm{vs.}}
\newc{\cl}{\rm {CL}}

\newc{\ev}{\ensuremath{\,\mathrm{eV}}}
\newc{\kev}{\ensuremath{\,\mathrm{keV}}}
\newc{\mev}{\ensuremath{\,\mathrm{MeV}}}
\newc{\gev}{\ensuremath{\,\mathrm{GeV}}}
\newc{\tev}{\ensuremath{\,\mathrm{TeV}}}
\newc{\MeV}{\mev} 
\newc{\TeV}{\tev}
\newc{\invpb}{\ensuremath{/\text{pb}}}
\newc{\invfb}{\ensuremath{/\text{fb}}}
\newc\nb{\ensuremath{\,\mathrm{nb}}} \newc\pb{\ensuremath{\,\mathrm{pb}}} \newc\fb{\ensuremath{\,\mathrm{fb}}}
\newc\pc{\ensuremath{\,\mathrm{pc}}}
\newc\kpc{\ensuremath{\,\mathrm{kpc}}}
\newc\mpc{\ensuremath{\,\mathrm{Mpc}}}
\newc\ps{\ensuremath{\,\mathrm{ps}}} 
\newc\cmeter{\ensuremath{\,\mathrm{cm}}} 
\newc\meter{\ensuremath{\,\mathrm{m}}} 
\newc\kmeter{\ensuremath{\,\mathrm{km}}}
\newc\second{\ensuremath{\,\mathrm{s}}}
\newc\msecond{\ensuremath{\,\mathrm{ms}}}
\newc\nsecond{\ensuremath{\,\mathrm{ns}}}
\newc\psecond{\ensuremath{\,\mathrm{ps}}}

\newc{\chisqmin}{\ensuremath{\chi^2_{\mathrm{min}}}}
\newc{\Delchisq}{\ensuremath{\Delta\chi^2}}
\newc{\chisq}{\ensuremath{\chi^2}}
\newc{\like}{\ensuremath{\mathcal{L}}}

\newc\lsim{\ensuremath{\mathrel{\rlap{\lower4pt\hbox{\hskip1pt$\sim$}}\raise1pt\hbox{$<$}}}}
\newc\gsim{\ensuremath{\mathrel{\rlap{\lower4pt\hbox{\hskip1pt$\sim$}}\raise1pt\hbox{$>$}}}}
\newc{\VEV}[1]{\ensuremath{\langle #1 \rangle}}
\newc{\dl}{\ensuremath{\stackrel{\leftarrow}{D}}}
\newc{\dr}{\ensuremath{\stackrel{\rightarrow}{D}}}

\newc{\bcenter}{\begin{center}}    \newc{\ecenter}{\end{center}}
\newc{\bfl}{\begin{flushleft}}    \newc{\efl}{\end{flushleft}}
\newc{\bfr}{\begin{flushright}}    \newc{\efr}{\end{flushright}}

\newc{\bi}{\begin{itemize}}
\newc{\ei}{\end{itemize}}
\newc{\bed}{\begin{description}}
\newc{\eed}{\end{description}}
\newc{\ben}{\begin{enumerate}}
\newc{\een}{\end{enumerate}}

\newc{\be}{\begin{equation}}
\newc{\ee}{\end{equation}}
\newc{\bea}{\begin{eqnarray}}
\newc{\eea}{\end{eqnarray}}
\newc{\ra}{\rightarrow}

\newc{\alphas}{\ensuremath{\alpha_s}}
\newc{\alphatwo}{\ensuremath{\alpha_2}}
\newc{\alphaone}{\ensuremath{\alpha_1}}
\newc{\alphai}[1]{\ensuremath{\alpha_{#1}}}
\newc{\alphaem}{\ensuremath{\alpha_{\mathrm{em}}}}
\newc{\alphaeff}{\ensuremath{\alpha_{\mathrm{eff}}}}
\newc{\sineff}{\ensuremath{\sin \theta_{\mathrm{eff}}}}
\newc{\sinsqeff}{\ensuremath{\sin^2 \theta_{\mathrm{eff}}}}
\newc{\dalphahad}{\ensuremath{\Delta \alpha_{\mathrm{had}}}}
\newc{\yt}{\ensuremath{h_t}} \newc{\yb}{\ensuremath{h_b}} \newc{\ytau}{\ensuremath{h_{\tau}}}
\newc\mz{\ensuremath{M_Z}} 
\newc\mw{\ensuremath{m_W}}
\newc\mZ{\mz}        \newc\mW{\mw}
\newc\mhsm{\ensuremath{ m_{H_{\mathrm{SM}}}}}
\newc{\mtop}{\ensuremath{ m_t}}               \newc{\mtpole}{\ensuremath{ M_t}}
\newc{\mbottom}{\ensuremath{ m_b}} 
\newc{\mtau}{\ensuremath{ m_{\tau}}}
\newc{\mt}{\mtpole}
\newc{\mb}{\mbottom} 
\newc{\rtwogg}{\ensuremath{R_{h_2}(\gamma\gamma)}}
\newc{\rtwozz}{\ensuremath{R_{h_2}(ZZ)}}
\newc{\ronegg}{\ensuremath{R_{h_1}(\gamma\gamma)}}
\newc{\ronezz}{\ensuremath{R_{h_1}(ZZ)}}
\newc{\rsiggg}{\ensuremath{R_{h_\textrm{sig}}(\gamma\gamma)}}
\newc{\rsigzz}{\ensuremath{R_{h_\textrm{sig}}(ZZ)}}
\newc{\llbar}{\ensuremath{\ell\bar{\ell}}}
\newc{\tauptaum}{\ensuremath{ \tau^+\tau^-}}
\newc{\qqbar}{\ensuremath{ q\bar{q}}} \newc{\ppbar}{\ensuremath{ p\bar{p}}}
\newc{\bbbar}{\ensuremath{ b\bar{b}}} \newc{\ttbar}{\ensuremath{ t\bar{t}}}
\newc{\ffbar}{\ensuremath{ f\bar{f}}} \newc{\tautaubar}{\ensuremath{ \tau\bar{\tau}}}
\newc{\mchi}{\ensuremath{m_\neutone}}
\newc{\squark}{\ensuremath{\tilde{q}}}
\newc{\slepton}{\ensuremath{\tilde{\ell}}}
\newc{\gluino}{\ensuremath{\tilde{g}}} 
\newc{\mgluino}{\ensuremath{{m_{\gluino}}}}

\newc{\sthw}{\ensuremath{ \sin\theta_W}}              \newc{\cthw}{\ensuremath{\cos\theta_W}}
\newc{\tanthw}{\ensuremath{ \tan\theta_W}}              \newc{\cotthw}{\ensuremath{\cot\theta_W}}
\newc{\ssqthw}{\ensuremath{\sin^2 \theta_W}}
\newc{\msbar}{\ensuremath{\overline{MS}}} \newc{\drbar}{\ensuremath{\overline{DR}}}
\newc{\mtmtsmmsbar}{\ensuremath{ m_t(m_t)^{\msbar}_{{\mathrm{SM}}}}}
\newc{\mtmtsmdrbar}{\ensuremath{ m_t(m_t)^{\drbar}_{{\mathrm{SM}}}}}
\newc{\mtmtmssmdrbar}{\ensuremath{ m_t(m_t)^{\drbar}_{{\mathrm{SUSY}}}}}
\newc{\mbmbmsbar}{\ensuremath{ m_b(m_b)^{\msbar} }}
\newc{\mbmbsmmsbar}{\ensuremath{ m_b(m_b)^{\msbar}_{{\mathrm{SM}}}}}
\newc{\mbmzsmmsbar}{\ensuremath{ m_b(\mz)^{\msbar}_{{\mathrm{SM}}}}}
\newc{\mbmzsmdrbar}{\ensuremath{ m_b(\mz)^{\drbar}_{{\mathrm{SM}}}}}
\newc{\mbmzmssmdrbar}{\ensuremath{ m_b(\mz)^{\drbar}_{{\mathrm{SUSY}}}}}
\newc{\mtaumzsmmsbar}{\ensuremath{ m_{\tau}(\mz)^{\msbar}_{{\mathrm{SM}}}}}
\newc{\mtaumzsmdrbar}{\ensuremath{ m_{\tau}(\mz)^{\drbar}_{{\mathrm{SM}}}}}
\newc{\mtaumzmssmdrbar}{\ensuremath{ m_{\tau}(\mz)^{\drbar}_{{\mathrm{SUSY}}}}}
\newc{\alphasmzms}{\ensuremath{\alpha_s(M_Z)^{\overline{MS}}}}
\newc{\alphaimzms}[1]{\ensuremath{\alpha_{#1}(M_Z)^{\overline{MS}}}}
\newc{\alphaemmz}{\ensuremath{\alpha_{\mathrm{em}}(M_Z)^{\overline{MS}}}}

\newc{\mzero}{\ensuremath{{m_0}}}
\newc{\mhalf}{\ensuremath{ m_{1/2}}}
\newc{\tanb}{\ensuremath{\tan\beta}}
\newc{\azero}{\ensuremath{ A_0}}
\newc{\sgnmu}{\ensuremath{\textrm{sgn}\,\mu}}
\newc{\atau}{\ensuremath{{A_{\tau}}}}
\newc{\mueff}{\ensuremath{\mu_{\rm{eff}}}}
\newc{\lam}{\ensuremath{{\lambda}}}
\newc{\kap}{\ensuremath{{\kappa}}}
\newc{\alam}{\ensuremath{{A_{\lambda}}}}
\newc{\akap}{\ensuremath{{A_{\kappa}}}}
\newc{\hs}{\ensuremath{ H_s}}      
\newc{\mhs}{\ensuremath{ m_{H_s}}} 
\newc{\mgut}{\ensuremath{ M_{\rm GUT}}}
\newc{\mplanck}{\ensuremath{ M_{\rm P}}}      \newc{\mpl}{\ensuremath{ M_{\rm Pl}}}
\newc{\msusy}{\ensuremath{ M_{\rm SUSY}}}      \newc{\ms}{\ensuremath{ M_{\rm S}}}
 \newc{\hu}{\ensuremath{ H_u}}       \newc{\hd}{\ensuremath{ H_d}}
 \newc{\mhu}{\ensuremath{ m_{H_u}}}       \newc{\mhd}{\ensuremath{ m_{H_d}}}
 \newc{\mhuew}{\ensuremath{ m^{\ast}_{H_u}}}       \newc{\mhdew}{\ensuremath{ m^{\ast}_{H_d}}}
 \newc{\mhuewsq}{\ensuremath{ m^{\ast\, 2}_{H_u}}}       \newc{\mhdewsq}{\ensuremath{ m^{\ast\, 2}_{H_d}}}
 \newc{\mhl}{\ensuremath{m_\hl}} 
 \newc{\mhone}{\ensuremath{m_{h_1}}} 
 \newc{\mhtwo}{\ensuremath{m_{h_2}}} 
 \newc{\mglu}{\ensuremath{m_{\tilde g}}} 
 \newc{\mul}{\ensuremath{m_{\tilde{u}_L}}} 
 \newc{\mstopone}{\ensuremath{m_{\tilde{t}_1}}} 
 \newc{\ma}{\ensuremath{m_A}} 
 \newc{\maone}{\ensuremath{m_{a_1}}} 
 \newc{\matwo}{\ensuremath{m_{a_2}}}
 \newc{\hone}{\ensuremath{h_1}}
 \newc{\htwo}{\ensuremath{h_2}}
 \newc{\aone}{\ensuremath{a_1}}
 \newc{\atwo}{\ensuremath{a_2}}

\newc{\sigsip}{\ensuremath{\sigma^{\rm SI}_{p}}}	\newc{\sigsin}{\ensuremath{\sigma^{\rm SI}_{n}}}
\newc{\sigsdp}{\ensuremath{\sigma^{\rm SD}_{p}}}	\newc{\sigsdn}{\ensuremath{\sigma^{\rm SD}_{n}}}
\newc{\sigsi}{\ensuremath{\sigma^{\rm SI}}}	\newc{\sigsd}{\ensuremath{\sigma^{\rm SD}}}
\newc{\abund}{\ensuremath{ \Omega h^2}}
\newc{\omegadm}{\ensuremath{ \Omega_{{\rm DM}}}}     \newc{\abunddm}{\ensuremath{ \Omega_{{\rm DM}} h^2}} 
\newc{\omegam}{\ensuremath{ \Omega_{{\rm m}}}}       \newc{\abundm}{\ensuremath{ \Omega_{{\rm m}} h^2}}
\newc{\omegab}{\ensuremath{ \Omega_{{\rm b}}}}	\newc{\abundb}{\ensuremath{ \Omega_{{\rm b}} h^2}}
\newc{\omegatot}{\ensuremath{ \Omega_{{\rm TOT}}}}
\newc{\omegacdm}{\ensuremath{ \Omega_{{\rm CDM}}}}   \newc{\abundcdm}{\ensuremath{ \Omega_{{\rm CDM}} h^2}}
\newc{\omegalambda}{\ensuremath{ \Omega_{\Lambda}}} \newc{\abundlambda}{\ensuremath{ \Omega_{\Lambda} h^2}}
\newc{\omegarad}{\ensuremath{ \Omega_{{\rm rad}}}}  \newc{\abundrad}{\ensuremath{ \Omega_{{\rm rad}} h^2}}
\newc{\rhocrit}{\ensuremath{ \rho_{\rm crit}}}
\newc{\rhochi}{\ensuremath{ \rho_{\chi}}}
\newc{\abunchi}{\ensuremath{\Omega_\chi h^2}}
\newc{\abundlsp}{\ensuremath{\Omega_{\rm LSP}h^2}}
\newcommand*{\abundchi}{\ensuremath{\Omega_\chi h^2}}

\newc{\amu}{\ensuremath{ a_{\mu}}}        \newc{\amususy}{\ensuremath{ a_{\mu}^{\mathrm{SUSY}}}}
\newc{\amuexpt}{\ensuremath{ a_{\mu}^{\mathrm{expt}}}}        \newc{\amusm}{\ensuremath{ a_{\mu}^{\mathrm{SM}}}}
\newc\deltaamu{\ensuremath{\Delta a_{\mu}}} \newc{\deltaamususy}{\ensuremath{\delta a_{\mu}^{\mathrm{SUSY}}}}
\newc\gmtwo{\ensuremath{ (g-2)_{\mu}}} 
\newc{\deltagmtwomususy}{\ensuremath{\delta\left(g-2\right)_{\mu}^{\mathrm{SUSY}}}}
\newc{\deltagmtwomu}{\ensuremath{\delta\left(g-2\right)_{\mu}}}
\newc\BR{\ensuremath{\textrm{BR}}}
\newc\bsgamma{\ensuremath{ b\rightarrow s \gamma }}
\newc\bxsgamma{\ensuremath{\overline{B}\rightarrow X_{s}\gamma}}
\newc\brbsgamma{\ensuremath{\BR\left(\bsgamma\right)}}
\newc\brbxsgamma{\ensuremath{\BR\left(\bxsgamma\right)}}
\newc\bsmumu{\ensuremath{B_s\to\mu^+\mu^-}}
\newc\brbsmumu{\ensuremath{\BR\left(B_s\to\mu^+\mu^-\right)}}
\newc\bdmmumu{\ensuremath{\overline{B}_d\to\mu^+\mu^-}}
\newc\bbbarmix{\ensuremath{\overline{B}_s\mbox{-}B_s}}      
\newc\delmbs{\ensuremath{\Delta M_{B_s}}}
\newc{\butaunu}{\ensuremath{B_u \rightarrow \tau \nu}}
\newc{\brbutaunu}{\ensuremath{\BR\left(B_u \rightarrow \tau \nu\right)}}

\newcommand*{\reftable}[1]{Table~\ref{#1}}         
       
\newcommand*{\reffig}[1]{Fig.~\ref{#1}}
 
        \newcommand*{\refeq}[1]{Eq.~\ref{#1}}
     \newcommand*{\refsec}[1]{Sec.~\ref{#1}}

\newcommand*{\neutone}{\ensuremath{\chi}}

\newcommand*{\stauc}{\text{\stau-coannihilation}}

\let\oldcite\cite
\renewcommand*{\cite}{~\oldcite}
\newcommand*{\datalr}{d}
\newcommand*{\hl}{\ensuremath{h}}

\newcommand*{\ha}{\ensuremath{A}}

\newcommand*{\mh}{\ensuremath{m_h}}


%% file: GoldenDecay.tex
\begin{fmffile}{GoldenDecay}
\begin{fmfgraph*}(200,100)
        \fmfleft{i1,i2,i3,i4}
        
        \fmfright{o1,o2,o3,o4}
        
        \fmf{phantom}{i1,v1,v2,v3,o1}
        \fmffreeze     
        \fmf{phantom}{i4,v4,v5,v6,o2}
        \fmffreeze    
        
        \fmf{scalar, label=$\s{q}_L$}{i1,v1}
        
        \fmf{fermion, label=$q$}{v1,v4}
        \fmf{phantom}{v4,o4}
        
        \fmf{gaugino, label=$\neut{2}$}{v1,v2}
        
        \fmf{fermion, label=$\ell^\mp$, label.side=left}{v2,v5}
        \fmf{phantom}{v5,o3}
        \fmf{scalar, label=$\s{\ell}_L$, label.side=left}{v3,v2}
        
        \fmf{fermion, label=$\ell^\pm$, label.side=right}{v6,v3}
        \fmf{phantom}{v6,o2}
        \fmf{gaugino, label=$\chi$}{v3,o1}
\end{fmfgraph*}
\end{fmffile}